\documentclass[%
aps,
twocolumn,
superscriptaddress,
nofootinbib,
 amsmath,amssymb,
prd,
nopacs,
longbibliography
]{revtex4-1}

\usepackage{graphicx}
\usepackage{dcolumn}
\usepackage{bm}
\usepackage{bbm}
\usepackage{xcolor}
\usepackage{upgreek}
\usepackage{amsmath}
\usepackage{indentfirst}
\usepackage{psfrag}
\usepackage{amssymb}
\usepackage{comment}
\usepackage{soul}
\usepackage[colorlinks,linkcolor=blue,citecolor=blue,urlcolor=blue,hyperindex,driverfallback=dvipdfm]{hyperref}
\usepackage[T1]{fontenc}
\allowdisplaybreaks
\def\app#1#2{%
  \mathrel{%
    \setbox0=\hbox{$#1\sim$}%
    \setbox2=\hbox{%
      \rlap{\hbox{$#1\propto$}}%
      \lower1.1\ht0\box0%
    }%
    \raise0.25\ht2\box2%
  }%
}

\definecolor{darkgreen}{rgb}{0.0, 0.4, 0.26}

\begin{document}

\title{Probing the electromagnetic response of dielectric antennas by vortex electron beams}

\author{Andrea Kone\v{c}n\'{a}}
\email{andrea.konecna@vutbr.cz}
\affiliation{Materials Physics Center, CSIC-UPV/EHU, 20018 Donostia-San Sebasti\'{a}n, Spain}
\affiliation{Central European Institute of Technology, Brno University of Technology, 612 00 Brno, Czech Republic}
\affiliation{Institute of Physical Engineering, Brno University of Technology, 616 69 Brno, Czech Republic}

\author{Miko\l{a}j K. Schmidt}
\affiliation{Materials Physics Center, CSIC-UPV/EHU, 20018 Donostia-San Sebasti\'{a}n, Spain}
\affiliation{{School of Mathematical and Physical Sciences}, Macquarie University, NSW 2109, Australia.}

\author{Rainer Hillenbrand}
\affiliation{CIC nanoGUNE BRTA and Department of Electricity and Electronics, UPV/EHU, 20018 Donostia-San Sebasti\'{a}n, Spain}
\affiliation{IKERBASQUE, Basque Foundation for Science, 48013 Bilbao, Spain}

\author{Javier Aizpurua}
\email{aizpurua@ehu.eus}
\affiliation{Materials Physics Center, CSIC-UPV/EHU, 20018 Donostia-San Sebasti\'{a}n, Spain}
\affiliation{Donostia International Physics Center DIPC, 20018 Donostia-San Sebasti\'{a}n, Spain}

\date{\today}

\begin{abstract}
    Focused beams of electrons, which act as both sources, and sensors of electric fields, can be used to characterise the electric response of complex photonic systems by locally probing the induced optical near fields. This functionality can be complemented by embracing the recently developed vortex electron beams (VEBs), made up of electrons with orbital angular momentum, which could, in addition, probe induced \textit{magnetic} near fields. 
    In this work, we revisit the theoretical description of this technique, dubbed vortex Electron Energy-Loss Spectroscopy (v-EELS). We map the fundamental, quantum-mechanical picture of the scattering of the VEB electrons to the intuitive classical models, which treat the electron beams as a superposition of linear electric and magnetic currents. We then apply this formalism to characterise the optical response of dielectric nanoantennas with v-EELS. Our calculations reveal that VEB electrons probe electric or magnetic modes with different efficiency, which can be adjusted by changing either beam vorticity or acceleration voltage to determine the nature of the probed excitations. We also study a chirally-arranged nanostructure, which in the interaction with electron vortices produces dichroism in electron energy loss spectra. Our theoretical work establishes VEBs as versatile probes that could provide information on optical excitations otherwise inaccessible with conventional electron beams.
\end{abstract}

\maketitle

\section{Introduction}
Electron energy-loss spectroscopy (EELS) in a scanning transmission electron microscope (STEM) \cite{egerton} is an emerging technique to characterize optical excitations with high spatial and spectral resolution \cite{GDA,Nelayah,lagos,ours}. Recent experimental and theoretical studies have demonstrated the capabilities of STEM-EELS to map near fields of localized surface polaritons in plasmonic and phononic nanostructures that are of high interest in the field of nanophotonics for their applications in focusing and engineering light below the diffraction limit \cite{giannini,pelton}. 

An alternative possibility to control light at the nanoscale is to use resonant electromagnetic (EM) modes in nanoparticles made of materials with high refractive index \cite{kuznetsov2016optically,Jahani,verre2018transition,Evlyukhin,albella_Surface-enahnced,cambiasso_GaP}, which have been, however, relatively rarely studied by near-field spectroscopic methods \cite{bakker,habteyes,miroshnichenko2015nonradiating,frolov,coenen_CL,groep_CL_dimers}. It has been shown only by recent experiments that focused electron beams such as those used in STEM-EELS can probe the response of dielectric  antennas \cite{crozier,kfir2019controlling,Alexander}. Here we explore the possibilities of using focused electron probes to distinguish electric and magnetic modes and \textit{hot spots} that are crucial for applications of dielectric particles in nanophotonics \cite{cihan2018silicon,regmi,rutckaia,Schmidt,vaskin}, and thus fully characterize the properties of their resonant modes. 

Interestingly, besides conventional electron beams, recent efforts have led to the generation of vortex electron beams (VEBs) in (S)TEM \cite{Bliokh_VEB,Lloyd_review,bliokh_review,Uchida,McMorran,Beche,Mafakheri,vanacore,tavabi}. VEBs carry orbital angular momentum (OAM), which could facilitate direct interaction of the beam with excitations of both electric and magnetic nature. Besides various applications in probing  magnetic fields \cite{Guzzinati,grillo2017observation}, magnetic transitions in bulk materials \cite{Verbeeck_2010,Lloyd,Yuan,Rusz} and chirality of crystals \cite{Juchtmans}, the introduction of VEBs (and other shaped beams) in electron microscopy by using adjustable phase plates \cite{Verbeeck2018,PAC,OFEM} might also open a pathway for symmetry-based selective excitation of EM modes in photonic nanostructures \cite{Ugarte, Guzzinati_2017, zanfrognini2018orbital}, separation of electric and magnetic modes \cite{Mohammadi}, or for developing the local investigation of the dichroic response of chiral nanoantennas \cite{Asenjo_garcia,zanfrognini2018orbital}.  

\begin{figure}
    \centering
    \includegraphics[width=\linewidth]{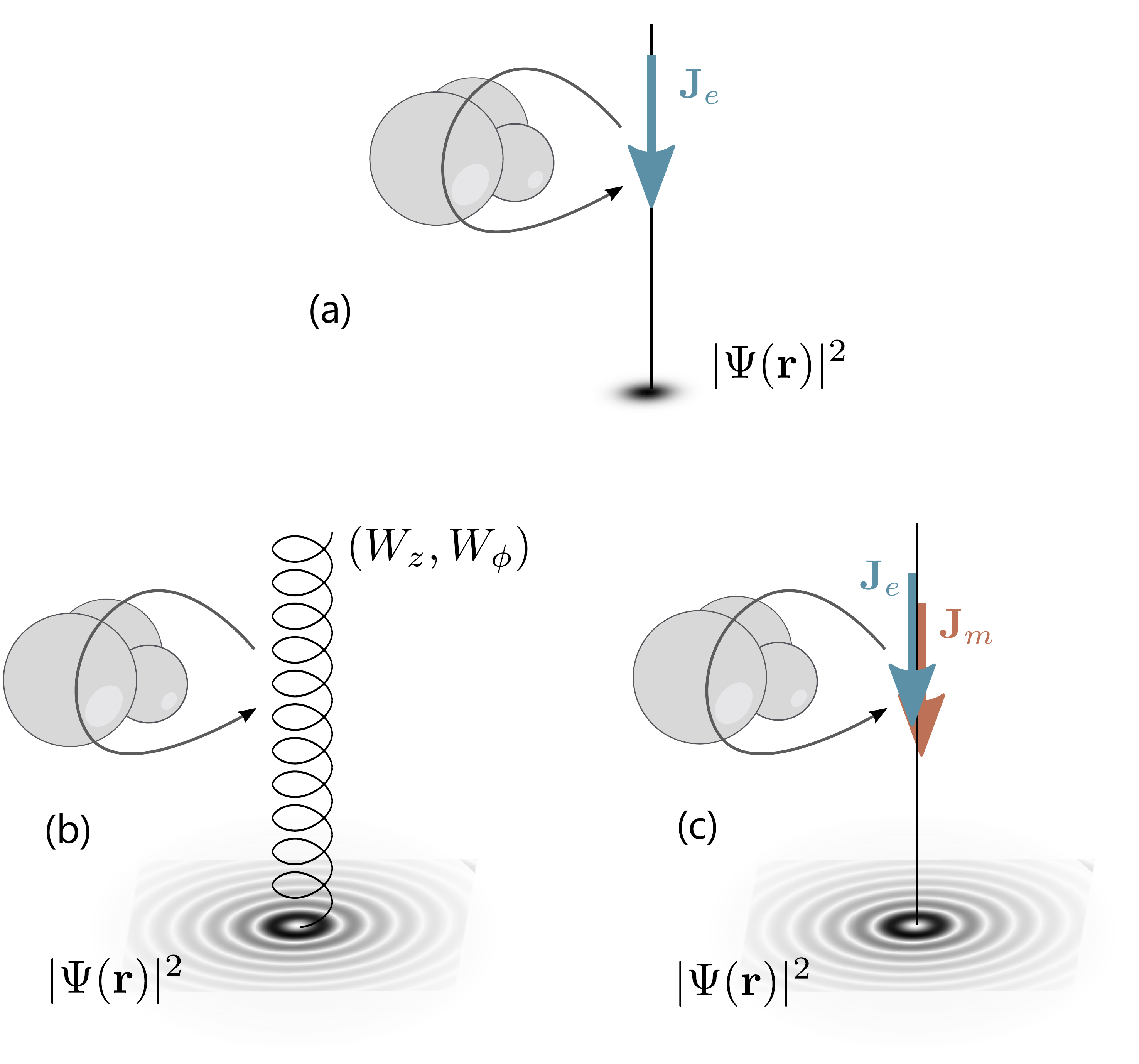}
    \caption{Models for describing the interaction between an electron beam with a sample. (a) In conventional STEM-EELS, the beam is modelled in the frequency domain as a broadband electric current density $\mathbf{J}_\mathrm{e}$, tightly localized in the transverse plane. (b,c) In v-EELS, the inelastic scattering of VEBs described by a structured wavefunction $\Psi(\mathbf{r})$ can be calculated by modelling the beam as a (b) helical electric current density characterized by axial and azimuthal components of the vector $\mathbf{W}$  (Section~\ref{sec:description}), or (c) a superposition of electric $\mathbf{J}_\mathrm{e}$ and magnetic $\mathbf{J}_\mathrm{m}$ currents, naturally extending the conventional STEM-EELS model (Section~\ref{sec:semi-classical}).}
    \label{fig:schematic}
\end{figure}

In this work, we show that STEM-EELS with the use of either a conventional or a vortex beam might be a suitable technique for distinguishing between the electric and magnetic nature of electromagnetic modes supported by dielectric antennas. We start by introducing a quantum-mechanical description of the inelastic interaction of VEBs with a general (classically responding) sample and with a point-like polarizable object, for which we obtain a closed form in the limit of a tightly focused VEB. To introduce the possibility of calculating the EEL spectrum with a VEB (v-EELS) for a spatially extended nanostructure, we find a source equivalent to the VEB within the framework of classical electrodynamics (see Fig.~\ref{fig:schematic}) and perform fully retarded calculations to retrieve the electromagnetic field arising from the VEB-sample interaction. We calculate EEL spectra considering the interaction with electron beams of both zero and non-zero OAM. We study single and dimer dielectric antennas of different shapes, particularly spherical and cylindrical structures made of silicon. We show that by varying excitation parameters or by comparing the spectra acquired with a non-vortex and a vortex beam, fast electrons preferentially couple to modes of electric and magnetic nature, respectively. Finally, we explore dichroism in v-EELS emerging for a chiral dielectric nanostructure.

\section{Theoretical framework for vortex electron energy loss spectroscopy at optical frequencies}
\subsection{Quantum-mechanical description of the beam}\label{sec:description}

The wavefunction of a vortex electron $\Psi$, can be described in the non-relativistic approximation (for discussion of the relativistic solutions, see Refs.~\cite{Bialynicki,Barnett}) as a solution of the Schr\"{o}dinger equation for a free-space moving electron with a non-vanishing OAM $l\hbar$. In cylindrical coordinates $(R,\phi,z)$, one of the simplest solutions 
takes the form of a Bessel beam \cite{Bliokh_VEB,Lloyd_review,bliokh_review} 
\begin{equation}
\Psi(R,\phi,z)=\frac{1}{\sqrt{ L}}\,\mathrm{e}^{\mathrm{i}q_z z}\underbrace{\frac{1}{\sqrt{A}}\mathrm{e}^{\mathrm{i}l\phi} J_l(Q R)}_{\psi_\perp},
\label{Eq:Bessel_state}
\end{equation}
where $Q$ and $q_z= m_\mathrm{e}v/\hbar$ are the radial and perpendicular wavevector components of the electron with mass $m_\mathrm{e}$ moving along the $z$ axis at velocity $v$, respectively, $A$ stands for a normalization area and $L$ for a normalization length, $\hbar$ is the reduced Planck constant. The Bessel function of order $l$,
$J_l(Q R)$, 
governs the radial shape variation of the beam profile, whereas the helical form of the wavefront is captured through the exponential term $\mathrm{e}^{\mathrm{i}l\phi}$.

We now express the probability of losing energy $\hbar\omega$ per electron considering a transition from a well-defined initial state $\Psi_\mathrm{i}$ to final states $\Psi_\mathrm{f}$ (following the formalism introduced in Ref.~\cite{Asenjo_garcia}), due to the interaction with the structured environment, as
\begin{align}
    \Gamma(\omega)&=\frac{2\hbar e^2 L}{\omega^2 m_\mathrm{e}^2 v}\sum_\mathrm{f}\int\mathrm{d}^3\mathbf{r}\,\mathrm{d}^3\mathbf{r}'\Psi_\mathrm{f}(\mathbf{r})\Psi^\ast_\mathrm{f}(\mathbf{r}')\nabla\left[\Psi_\mathrm{i}^\ast(\mathbf{r})\right]\cdot\nonumber\\
    &\cdot\mathrm{Im}\left[\hat{\mathbf{G}}(\mathbf{r},\mathbf{r}',\omega)\right]\cdot\nabla\left[\Psi_\mathrm{i}(\mathbf{r}')\right]\delta(\epsilon_\mathrm{f}-\epsilon_\mathrm{i}+\omega),
    \label{Eq:GamQ1}
\end{align}
where $e$ is the elementary charge, $\hbar\epsilon_\mathrm{f/i}$ is the final/initial electron energy, $\hat{\mathbf{G}}$ is the Green's tensor describing the electromagnetic response of the probed structure and where we sum over the final states.

In the following, we restrict ourselves to the states $\Psi_\mathrm{i}=\mathrm{e}^{\mathrm{i}q_{z,\mathrm{i}} z}\psi_{\perp,\mathrm{i}}/\sqrt{L}$ and $\Psi_\mathrm{f}=\mathrm{e}^{\mathrm{i}q_{z,\mathrm{f}} z}\psi_{\perp,\mathrm{f}}/\sqrt{L}$ with initial and final longitudinal wavevector components $q_{z,\mathrm{i}}$ and $q_{z,\mathrm{f}}$, respectively. We further consider $\psi_{\perp,\mathrm{f}}=\mathrm{e}^{\mathrm{i}l_\mathrm{f}\phi}J_{l_\mathrm{f}}(Q_\mathrm{f}R)/\sqrt{A}$ with a set of possible transverse wavevectors $Q_\mathrm{f}$ and a well-defined initial transverse wavefunction component $\psi_{\perp,\mathrm{i}}\approx 1/(Q_\mathrm{c,i}\sqrt{\pi})\int_0^{Q_\mathrm{c,i}} Q_\mathrm{i}\mathrm{d}Q_\mathrm{i}\,\mathrm{e}^{\mathrm{i}l_\mathrm{i}\phi}J_{l_\mathrm{i}}(Q_\mathrm{i}R)$ where $Q_\mathrm{c,i}$ is an initial wavevector cutoff. The centers of the forming and collection apertures are assumed to be aligned on top of each other.

Considering a detector imposing a cutoff of transverse final wavevectors $Q_\mathrm{c}$ due to a finite collection angle, we can replace the schematic sum over final states $\sum_{\rm f}$ with ${L A}/(4 \pi^2) \int \mathrm{d} q_{z, \mathrm{f}}  \int_0^{Q_\mathrm{c}} Q_\mathrm{f} \mathrm{d} Q_\mathrm{f}$. We also rewrite the spatial integrals over $\mathbf{r}$ and $\mathbf{r}'$ in cylindrical coordinates, and collect all the $z$- and $z'$-dependent exponentials from the $\Psi$'s [see Eq.~\eqref{Eq:Bessel_state}] to evaluate the integral in the non-recoil approximation 
\begin{equation}
    \int \mathrm{d} q_{z,\mathrm{f}} e^{\mathrm{i}(q_{z, \mathrm{f}}-q_{z,\mathrm{i}})(z-z')}\delta(\epsilon_\mathrm{f}-\epsilon_\mathrm{i}+\omega)=\frac{e^{-\mathrm{i}\omega(z-z')/v}}{v}.
\end{equation} 
Defining 
\begin{align}
\hat{\bm{\mathcal{G}}}(\mathbf{R},\mathbf{R}',\omega)=\int\mathrm{d}z\,\mathrm{d}z'\,\mathrm{e}^{-\mathrm{i}\omega(z-z')/v}\hat{\mathbf{G}}(\mathbf{r},\mathbf{r}',\omega),   
\label{Eq:Gcal}
\end{align}
we can express the loss probability as
\begin{align}
    &\Gamma(\omega)=\frac{e^2}{2\pi^3 \hbar\omega^2Q_\mathrm{c,i}^2}\int_0^{{Q_\mathrm{c}}} \,Q_\mathrm{f}\mathrm{d}Q_\mathrm{f}\mathrm{Im}\left[\int R\mathrm{d}R\,R'\mathrm{d}R'\right.\nonumber\\ 
    &\left. J_{l_\mathrm{f}}(Q_\mathrm{f}R) J_{l_\mathrm{f}}(Q_\mathrm{f}R')f(R)f(R')\int_0^{2\pi}\mathrm{d}\phi\,\mathrm{d}\phi'\,\mathrm{e}^{\mathrm{i}\Delta l(\phi-\phi')}\right.\nonumber\\
    &\left.\mathbf{V}^\ast(R,\phi)\cdot\hat{\bm{\mathcal{G}}}(R,R',\phi,\phi',\omega)\cdot\mathbf{V}(R',\phi')\right].\label{Eq:GamQ}
\end{align}
Above we introduced $f(R)=\int_0^{Q_\mathrm{c,i}}Q_\mathrm{i}\mathrm{d}Q_\mathrm{i}J_{l_\mathrm{i}}(Q_\mathrm{i}R)$, $\Delta l=l_\mathrm{f}-l_\mathrm{i}$, and defined the vector field $\mathbf{V}$ related to the gradient of electron's wavefunctions
\begin{align}
    \mathbf{V}(R,\phi)=\mathbf{e}_z+\frac{\mathbf{e}_\phi l_\mathrm{i}}{R q_{z,\mathrm{i}}}-\frac{\mathbf{e}_R \mathrm{i} f'(R)}{q_{z,\mathrm{i}}f(R)},\label{Eq:V}
\end{align}
where $\mathbf{e}_i$ denote unit vectors along the directions $i$.

\begin{figure}
    \centering
    \includegraphics[width=\linewidth]{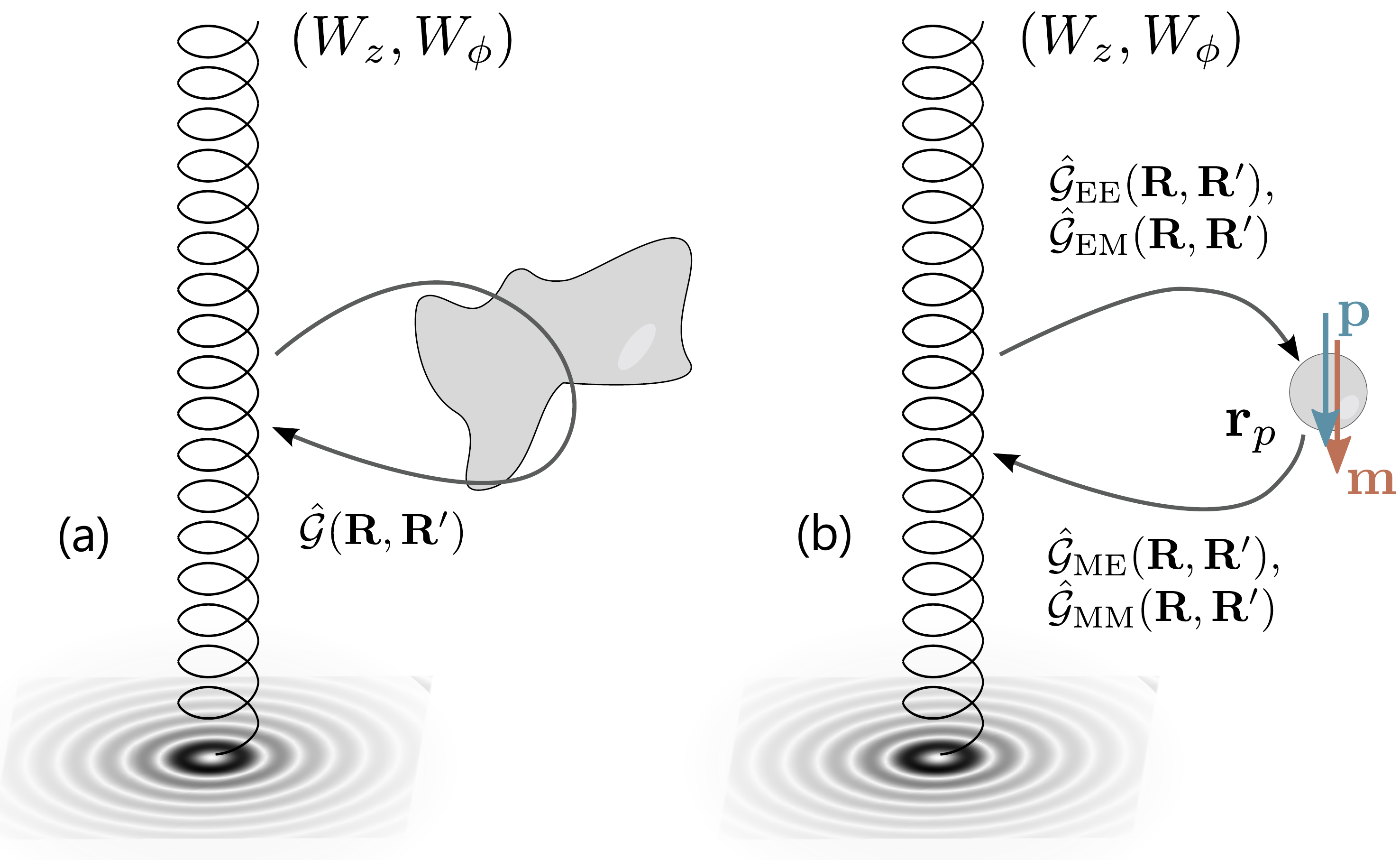}
    \caption{Illustration of the semi-classical framework for calculating the interaction of the VEB with an environment. In (a) the electron beam is modelled as quantum current densities with axial and azimuthal components characterized by the vector $\mathbf{W}$, acting like quantum analogues of the electric currents. Its interaction with the environment is dictated by the Green's function $\hat{\bm{\mathcal{G}}}$ [see Eq.~\eqref{Eq:GamQ}] which describes the scattering of radiation from the electric dipolar source in the environment. In (b) we consider an example of such interaction, with the environment modelled as a dipolar point-like scatterer at $\mathbf{r}_\mathrm{p}$, and axial polarizabilities $\hat{\alpha}_\text{EE}$, $\hat{\alpha}_\text{EM}$, $\hat{\alpha}_\text{ME}$, and $\hat{\alpha}_\text{MM}$. The electric current density created electric and magnetic fields at the position of the scatterer, according to the Green's function decomposed as in Eq.~\eqref{Eq:Gdecompose}, and induces dipolar momenta $\mathbf{p}$ and $\mathbf{m}$, respectively.}
    \label{fig:FigS1}
\end{figure}

When all electrons are collected by the detector ($Q_\mathrm{c}\rightarrow\infty$), we can use the identity $\int_0^\infty x\, \mathrm{d}x\, J_l(x R)J_l(x R')=\delta(R-R')/R'$ to perform the integral over the final transverse wavevectors to get
\begin{align}
    &\Gamma(\omega)=\frac{e^2}{2\pi^3 \hbar\omega^2Q_\mathrm{c,i}^2}\mathrm{Im}\left[\int_0^\infty R\,\mathrm{d}R\,f^2(R)\int_0^{2\pi} \mathrm{d}\phi\,\mathrm{d}\phi'\right.\nonumber\\
    &\left.\mathrm{e}^{\mathrm{i}\Delta l(\phi-\phi')}\, \mathbf{V}^\ast(R,\phi)\cdot\hat{\bm{\mathcal{G}}}(R,R,\phi,\phi',\omega)\cdot\mathbf{V}(R,\phi')\right].\label{Eq:GamQ2}
\end{align}
Note that $f^2(R)$ will be strongly peaked around an effective initial VEB radius $R_{0,l_\mathrm{i}}$ given by the initial cutoff value. Therefore if we consider $\hat{\bm{\mathcal{G}}}$ slowly varying around $R_{0,l_\mathrm{i}}$, we can roughly approximate the integral over $R$ by $ \int_0^\infty R\, \mathrm{d}R\,f^2(R)= Q_\mathrm{c,i}^2/2$ to obtain
\begin{align}
    \Gamma(\omega)&\approx \frac{e^2}{4\pi^3 \hbar\omega^2}\mathrm{Im}\left[\int_0^{2\pi}
    \mathrm{d}\phi\,\mathrm{d}\phi'\, \mathrm{e}^{\mathrm{i}\Delta l(\phi-\phi')}\right.\nonumber\\
    &\left.\mathbf{W}^\ast(R_{0,l_\mathrm{i}},\phi)\cdot\,\hat{\bm{\mathcal{G}}}(R_{0,l_\mathrm{i}},R_{0,l_\mathrm{i}},\phi,\phi',\omega)\cdot\mathbf{W}(R_{0,l_\mathrm{i}},\phi')\right]\label{Eq:Gamapprox},
\end{align}
where we disregarded the radial component of $\mathbf{V}$ as it is much smaller than the other components around $R_{0,l_\mathrm{i}}$, yielding
\begin{align}
    &\mathbf{W}(R,\phi)=\mathbf{e}_z+ \frac{\mathbf{e}_\phi l_\mathrm{i}}{R_{0,l_\mathrm{i}} q_{z,\mathrm{i}}}\label{Eq:Vappr}.
\end{align}
This formulation of v-EELS is depicted schematically in Fig.~\ref{fig:FigS1}(a).

\subsection{Loss probability for a VEB interacting with a point-like dipolar particle in a focused beam limit}
\label{Sec:dipole}
The formulation of the loss probability given in Eq.~\eqref{Eq:Gamapprox} is general, but does not easily simplify to a classical picture, widely embraced to address conventional EELS. To aid that simplification, here we present a calculation of loss probability due to the interaction with a specific system --- a point-like dipolar particle shown in Fig.~\ref{fig:FigS1}(b). We will use this result in the following Section~\ref{sec:semi-classical} to identify a semi-classical description, shown schematically in Fig.~\ref{fig:schematic}(c).

The point-like scatterer is situated at $\mathbf{r}_\mathrm{p}=(\mathbf{R}_\mathrm{p},z_\mathrm{p})$, and is
electrically and magnetically polarizable in the $z$ direction, and 
characterized by the electric polarizability tensor $\hat{\mathbf{\alpha}}_\mathrm{EE}=\alpha_\mathrm{EE}(\mathbf{0},\mathbf{0},\mathbf{e}_z)$, magnetic polarizability tensor $\hat{\mathbf{\alpha}}_\mathrm{MM}=\alpha_\mathrm{MM}(\mathbf{0},\mathbf{0},\mathbf{e}_z)$, and crossed electric-magnetic and magnetic-electric polarizability tensors $\hat{\mathbf{\alpha}}_\mathrm{EM}=\alpha_\mathrm{EM}(\mathbf{0},\mathbf{0},\mathbf{e}_z)$ and $\hat{\mathbf{\alpha}}_\mathrm{ME}=\alpha_\mathrm{ME}(\mathbf{0},\mathbf{0},\mathbf{e}_z)$, respectively. The latter two tensors are responsible for a dichroic response and fulfill $\hat{\mathbf{\alpha}}_\mathrm{ME}=-\hat{\mathbf{\alpha}}_\mathrm{EM}^\mathrm{T}$. For simplicity, we assume the response only in the $z$ direction, however, a general direction should be considered in a realistic scenario. The Green's tensor is then
\begin{align}
&\hat{\mathbf{G}}(\mathbf{r},\mathbf{r}',\omega)=\nonumber\\
&=\sum\limits_{\substack{i= \\ \lbrace \mathrm{E,M}\rbrace}}\sum\limits_{\substack{j= \\ \lbrace \mathrm{E,M}\rbrace}}\hat{\mathbf{G}}_{\mathrm{E}i}(\mathbf{r}-\mathbf{r}_\mathrm{p},\omega)\cdot\hat{\bm{\alpha}}_{ij}(\omega)\cdot\hat{\mathbf{G}}_{j\mathrm{E}}(\mathbf{r}_\mathrm{p}-\mathbf{r}',\omega)
\end{align}
with $\hat{\mathbf{G}}_\mathrm{EE}(\mathbf{r})=c^2\hat{\mathbf{G}}_\mathrm{MM}(\mathbf{r})=(k^2\hat{\mathbf{I}}+\nabla\otimes\nabla)\,\mathrm{e}^{\mathrm{i}k r}/(4\pi\varepsilon_0 r)$ and $\hat{\mathbf{G}}_\mathrm{EM}(\mathbf{r})=-\hat{\mathbf{G}}_\mathrm{ME}(\mathbf{r})=-\nabla\times\hat{\mathbf{G}}_\mathrm{EE}(\mathbf{r})/(\mathrm{i}kc)$, where $k=\omega/c$ with the speed of light $c$, $\varepsilon_0$ is the vacuum permittivity and $\otimes$ denotes tensor product.

The integrals over $z$ (and $z'$) involved in $\hat{\bm{\mathcal{G}}}$ [see Eq.~\eqref{Eq:Gcal}] are analytical \cite{Asenjo_garcia}:
\begin{align}
    &\hat{\bm{\mathcal{G}}}_\mathrm{EE}=\int\mathrm{d}z\,\mathrm{e}^{-\mathrm{i}\omega z/v}\hat{\mathbf{G}}_\mathrm{EE}(\mathbf{r}-\mathbf{r}_\mathrm{p})\nonumber\\
    &=\frac{1}{2\pi\varepsilon_0}(k^2\hat{\mathbf{I}}+\nabla_{\mathbf{r}_\mathrm{p}}\otimes\nabla_{\mathbf{r}_\mathrm{p}})\mathrm{e}^{-\mathrm{i}\omega z_\mathrm{p}/v}K_0\left(\frac{\omega\lvert\mathbf{R}-\mathbf{R}_\mathrm{p}\rvert^2}{v\gamma}\right),\label{Eq:green}
\end{align}
and 
\begin{align}
    \hat{\bm{\mathcal{G}}}_\mathrm{EM}&=\int\mathrm{d}z\,\mathrm{e}^{-\mathrm{i}\omega z/v}\hat{\mathbf{G}}_\mathrm{EM}(\mathbf{r}-\mathbf{r}_\mathrm{p})=\frac{1}{\mathrm{i}k c}\nabla_{\mathbf{r}_\mathrm{p}}\times\hat{\bm{\mathcal{G}}}_\mathrm{EE},
\end{align}
which we use to obtain
\begin{align}
    \hat{\bm{\mathcal{G}}}(\mathbf{R},\mathbf{R}')&=\sum\limits_{\substack{i= \\ \lbrace \mathrm{E,M}\rbrace}}\sum\limits_{\substack{j= \\ \lbrace \mathrm{E,M}\rbrace}} \hat{\bm{\mathcal{G}}}_{\mathrm{E}i}(\mathbf{R},\mathbf{R}_\mathrm{p})\cdot\hat{\bm{\alpha}}_\mathrm{ij}\cdot \hat{\bm{\mathcal{G}}}_{j\mathrm{E}}(\mathbf{R}_\mathrm{p},\mathbf{R}'),\label{Eq:Gdecompose}\nonumber
\end{align}
where all tensors also depend on $\omega$.
We can expand the products of the currents and Green's functions in the rhs of Eq.~\eqref{Eq:Gamapprox} as
\begin{align}
    \mathbf{W}^* \cdot \hat{\bm{\mathcal{G}}} \cdot \mathbf{W} = &~W_z^* \mathcal{G}_{zz} W_z + W_z^* \mathcal{G}_{z\phi} W_\phi\nonumber \\
    &+W_\phi^* \mathcal{G}_{\phi z} W_z + W_\phi^* \mathcal{G}_{\phi \phi} W_\phi.
\end{align}
The first term expands as 
\begin{align}
    W_z^* \mathcal{G}_{zz} W_z = ~W_z^* \Big[&(\hat{\bm{\mathcal{G}}}_{\mathrm{EE}} \cdot \hat{\alpha}_{\mathrm{EE}} \cdot \hat{\bm{\mathcal{G}}}_{\mathrm{EE}})_{zz}\nonumber \\
    &+\underbrace{(\hat{\bm{\mathcal{G}}}_{\mathrm{EM}} \cdot \hat{\alpha}_{\mathrm{ME}} \cdot \hat{\bm{\mathcal{G}}}_{\mathrm{EE}})_{zz}}_0\nonumber \\
    &+\underbrace{(\hat{\bm{\mathcal{G}}}_{\mathrm{EE}} \cdot \hat{\alpha}_{\mathrm{EM}} \cdot \hat{\bm{\mathcal{G}}}_{\mathrm{ME}})_{zz}}_0\nonumber \\    &+\underbrace{(\hat{\bm{\mathcal{G}}}_{\mathrm{EM}} \cdot \hat{\alpha}_{\mathrm{MM}} \cdot \hat{\bm{\mathcal{G}}}_{\mathrm{ME}})_{zz}}_0\Big] W_z,
\end{align}
where the three terms vanish due to the symmetries of the Green's functions (e.g. $(\hat{\bm{\mathcal{G}}}_\mathrm{ME})_{zz}=(\hat{\bm{\mathcal{G}}}_\mathrm{EM})_{zz}=0$), and the axial form of the polarizability. Similarly, one we can simplify
\begin{align}
    W_z^* \mathcal{G}_{z\phi} W_\phi = ~W_z^* (\hat{\bm{\mathcal{G}}}_{\mathrm{EE}} \cdot \hat{\alpha}_{\mathrm{EM}} \cdot \hat{\bm{\mathcal{G}}}_{\mathrm{ME}})_{z\phi} W_\phi,
\end{align}
\begin{align}
    W_\phi^* \mathcal{G}_{\phi z} W_z = ~W_\phi^* (\hat{\bm{\mathcal{G}}}_{\mathrm{EM}} \cdot \hat{\alpha}_{\mathrm{ME}} \cdot \hat{\bm{\mathcal{G}}}_{\mathrm{EE}})_{\phi z} W_z,
\end{align}
\begin{align}
    W_\phi^* \mathcal{G}_{\phi z} W_\phi = ~W_\phi^* (\hat{\bm{\mathcal{G}}}_{\mathrm{EM}} \cdot \hat{\alpha}_{\mathrm{MM}} \cdot \hat{\bm{\mathcal{G}}}_{\mathrm{ME}})_{\phi \phi} W_\phi.
\end{align}

We now split the total loss probability as
\begin{align}
    \Gamma(\omega)=\Gamma_\mathrm{EE}(\omega)+\Gamma_\mathrm{EM}(\omega)+\Gamma_\mathrm{ME}(\omega)+\Gamma_\mathrm{MM}(\omega),\label{Eq:Eels_sum}
\end{align}
where the individual contributions are defined by the products of the components of $\mathbf{W}$, i.e. $\Gamma_\mathrm{EE}\propto |W_z|^2$, $\Gamma_\mathrm{EM}\propto W_z^* W_\phi$, $\Gamma_\mathrm{ME}\propto W_\phi^* W_z$, $\Gamma_\mathrm{MM}\propto |W_\phi|^2$:

\begin{align}
     &\Gamma_\mathrm{EE}(\omega)\approx\frac{e^2}{4\pi^3 \hbar\omega^2}
     \mathrm{Im}\left[\int_0^{2\pi} \mathrm{d}\phi\,\mathrm{d}\phi'\,\mathrm{e}^{\mathrm{i}\Delta l(\phi-\phi')}\right.\nonumber\\
     &\left.\hat{\mathcal{G}}_{zz'}(R_{0,l_\mathrm{i}},R_{0,l_\mathrm{i}},\phi,\phi',\omega)\right]\nonumber\\
     &=\frac{e^2\omega^2 \mathrm{Im}[\alpha_\mathrm{EE}(\omega)]}{4\pi^3\varepsilon_0^2 \hbar v^4\gamma^4} \left[I_{\Delta l}\left(\frac{\omega R_{0,l_\mathrm{i}}}{v\gamma}\right)K_{\Delta l}\left(\frac{\omega R_\mathrm{p}}{v\gamma}\right)\right]^2,
     \label{Eq:Gam_l0}
\end{align}
where we assumed $R_{0,l_\mathrm{i}}\leq R_\mathrm{p}$, and introduced the Lorentz factor $\gamma=1/\sqrt{1-v^2/c^2}$. With the same assumptions, we obtain:
\begin{align}
    &\Gamma_\mathrm{MM}(\omega)\approx \frac{e^2 l^2_\mathrm{i}}{4\pi^3 \hbar\omega^2 {R^2_{0,l_\mathrm{i}}q^2_{z,\mathrm{i}}}} \mathrm{Im}\left[\int_0^{2\pi} \mathrm{d}\phi\,\mathrm{d}\phi'\right.\nonumber\\
    &\left.\mathrm{e}^{\mathrm{i}\Delta l(\phi-\phi')}\hat{\mathcal{G}}_{\phi\phi'}(R_{0,l_\mathrm{i}},R_{0,l_\mathrm{i}},\phi,\phi',\omega)\right]\nonumber\\
     &=\frac{e^2 l_\mathrm{i}^2\omega^2\mathrm{Im}[\alpha_\mathrm{MM}(\omega)]}{4\pi^3\varepsilon_0^2 \hbar v^2c^4\gamma^2 R_{0,l_\mathrm{i}}^2q_{z,\mathrm{i}}^2}\left[I'_{\Delta l}\left(\frac{\omega R_{0,l_\mathrm{i}}}{v\gamma}\right)K_{\Delta l}\left(\frac{\omega R_\mathrm{p}}{v\gamma}\right)\right]^2,
\end{align}
and
\begin{align}
    &\Gamma_\mathrm{\lbrace EM/ME\rbrace}(\omega)\nonumber\\
    &\approx \frac{e^2 l_\mathrm{i}}{4\pi^3 \hbar\omega^2 {R_{0,l_\mathrm{i}}q_{z,\mathrm{i}}}} \mathrm{Im}\left[\int_0^{2\pi} \mathrm{d}\phi\,\mathrm{d}\phi'\mathrm{e}^{\mathrm{i}\Delta l(\phi-\phi')}\right.\nonumber\\\
    &\left.\left\lbrace\hat{\mathcal{G}}_{z\phi'}(R_{0,l_\mathrm{i}},R_{0,l_\mathrm{i}},\phi,\phi')/\hat{\mathcal{G}}_{\phi z'}(R_{0,l_\mathrm{i}},R_{0,l_\mathrm{i}},\phi,\phi')\right\rbrace\right]\nonumber\\
     &=\frac{\lbrace \mp\rbrace e^2 l_\mathrm{i}\omega^2\mathrm{Re}\left[\alpha_\mathrm{\lbrace EM/ME\rbrace}(\omega)\right]}{4\pi^3\epsilon_0^2 \hbar v^3 c^2\gamma^3 {R_{0,l_\mathrm{i}}q_{z,\mathrm{i}}}}\nonumber\\
     &\times I'_{\Delta l}\left(\frac{\omega R_{0,l_\mathrm{i}}}{v\gamma}\right)I_{\Delta l}\left(\frac{\omega R_{0,l_\mathrm{i}}}{v\gamma}\right)K^2_{\Delta l}\left(\frac{\omega R_\mathrm{p}}{v\gamma}\right).
\end{align}
We evaluated the integrals following Ref.~\cite{Asenjo_garcia}. We note that for a well-focused VEB with $\omega R_{0,l}/(v \gamma)\rightarrow 0$, $\Delta l=0$, and thus using $I_0(x)\sim 1$ and $I_1(x)\sim x/2$ for small arguments, we obtain
\begin{align}
    &\Gamma(\omega)=\frac{e^2 \omega^2}{4\pi^3\varepsilon_0^2\hbar v^4\gamma^4}K^2_0\left(\frac{\omega R_\mathrm{p}}{v\gamma}\right)\left\lbrace \mathrm{Im}[\alpha_\mathrm{EE}(\omega)]\right.\nonumber\\
    &\left.+\frac{ l_\mathrm{i}^2\omega^2\mathrm{Im}[\alpha_\mathrm{MM}(\omega)]}{4 c^4q_{z,\mathrm{i}}^2}- \frac{l_\mathrm{i}\omega}{2c^2q_{z,\mathrm{i}}}\mathrm{Re}\left[\alpha_\mathrm{EM}(\omega)-\alpha_\mathrm{ME}(\omega)\right]\right\rbrace.
    \label{Eq:Gamma_point_dipole_approx}
\end{align}

The well-focused-VEB limit is very accurate for $R_{0,l}\lesssim 10$~nm at optical frequencies and for typical TEM acceleration voltages. Such effective radii are achievable even for relatively large $l$ ($\sim 100$) as shown in Appendix~\ref{app:radius}. We also note that for $l_\mathrm{i}=0$, the result above coincides with the classical limit with a well-focused beam at the origin interacting with an electric dipole oriented along the $z$ axis.

The loss probability in Eq.~\eqref{Eq:Gamma_point_dipole_approx} can be readily evaluated for a beam carrying OAM $+\hbar l_\mathrm{i}$ and $-\hbar l_\mathrm{i}$, which yields the dichroic signal
\begin{align}
    \Gamma_{-\lvert l_\mathrm{i}\rvert}-\Gamma_{\lvert l_\mathrm{i}\rvert}=\frac{e^2 \lvert l_\mathrm{i}\rvert \omega^3}{2\pi^3\varepsilon_0^2\hbar v^4\gamma^4c^2q_{z,\mathrm{i}}}K^2_0\left(\frac{\omega R_\mathrm{p}}{v\gamma}\right) \mathrm{Re}\left[\alpha_\mathrm{EM}(\omega)\right],
\end{align}
where we used $\alpha_\mathrm{ME}=-\alpha_\mathrm{EM}$. Interestingly, the proportionality of the dichroic signal to $\mathrm{Re}[\alpha_\mathrm{EM}]$ holds also for the difference of optical absorption obtained with right- and left-handed circularly polarized light ($A_\mathrm{RCP}-A_\mathrm{LCP}$) \cite{tang}. However, compared to the optical circular dichroism, the local excitation by a focused VEB makes it possible to probe the dichroic response with high spatial resolution, which is manifested 
in the fast decay of the signal strength with increasing distance of the electron 
beam from the particle, as the EEL probability strongly depends 
on the field accompanying the VEBs.

Importantly, considering typical scaling of polarizability components ($\alpha_\mathrm{MM}\sim c^{2} \alpha_\mathrm{EE}$ and $\alpha_\mathrm{EM}\sim c\alpha_\mathrm{EE}$), we can see that the contribution to the 
loss probability due to the interaction with the electric dipole will be dominant as $\omega/(2q_{z,\mathrm{i}}c)\sim 10^{-5}$ at optical frequencies and typical velocities of electron probes. Significant improvements in detecting dichroism or purely magnetic response from point-like objects with well-focused VEBs could be achieved by employing slower electrons with large OAM.

\subsection{Semi-classical formalism}
\label{sec:semi-classical}
The interaction of a well-focused VEB with a sample can also be expressed using a semi-classical formalism by introducing effective frequency-dependent electric and magnetic line currents $\mathbf{J}_\mathrm{e}$ and $\mathbf{J}_\mathrm{m}$, respectively, representing the VEB. These sources induce the electromagnetic response of a sample, which acts back on the electron beam and causes its energy loss \cite{GDA} [see schematic in Fig.~\ref{fig:FigS2}(a)]. While we expect that the electric current of a VEB will have a simple form, identical to that used throughout the literature on conventional EELS \cite{GDA}, we seek to identify the exact expression for the magnetic current.

The total energy loss consists of the energy loss experienced by both the electric and the magnetic current components of the beam:  $\Delta E=\Delta E_\mathrm{e}+\Delta E_\mathrm{m}$. The electric and the magnetic energy losses in the non-recoil approximation are given by \cite{Mohammadi}
\begin{align}
\Delta E_\mathrm{e}&=\frac{-1}{\pi}\int\limits_0^\infty\mathrm{d}\omega\int\limits_{-\infty}^\infty\mathrm{d}\mathbf{r}~\mathrm{Re}\left[\mathbf{E}^\mathrm{ind}(\mathbf{r},\omega)\cdot\mathbf{J}^*_\mathrm{e}(\mathbf{r},\omega)\right],\label{eq:del}
\end{align}
\begin{align}
\Delta E_\mathrm{m}&=\frac{-1}{\pi}\int\limits_0^\infty\mathrm{d}\omega\int\limits_{-\infty}^\infty\mathrm{d}\mathbf{r}~\mathrm{Re}\left[\mathbf{B}^\mathrm{ind}(\mathbf{r},\omega)\cdot\mathbf{J}^*_\mathrm{m}(\mathbf{r},\omega)\right],\label{eq:dmag}
\end{align}
where $\mathbf{E}^\mathrm{ind}$ and $\mathbf{B}^\mathrm{ind}$ are electric and the magnetic fields, respectively, induced by the radiation from the electric and magnetic currents $\mathbf{J}_\mathrm{e}$ and $\mathbf{J}_\mathrm{m}$, respectively.
The last two expressions can be related using the invariance of the Maxwell's equations, and the corresponding Green's functions, in free space, under the transformation \cite{NH06}
\begin{equation}
\mathbf{E}\rightarrow c\mathbf{B}, \quad \mathbf{B}\rightarrow -\frac{\mathbf{E}}{c}, \quad \mathbf{p}\rightarrow \frac{\mathbf{m}}{c}.
\label{Eq:invariance_Maxwell}
\end{equation}

We can now introduce the electric and magnetic loss probabilities $\Gamma_\mathrm{e}$ and $\Gamma_\mathrm{m}$, respectively, as $\Delta E_\mathrm{\lbrace e/m\rbrace}=\int_0^\infty\mathrm{d}\omega\,\hbar\omega\,\Gamma_\mathrm{\lbrace e/m\rbrace}$. The total loss probability
\begin{align}
    \Gamma(\omega)=\Gamma_\mathrm{e}(\omega)+\Gamma_\mathrm{m}(\omega)
\end{align} 
then corresponds to the measured electron energy loss spectrum for the case that a perfectly focused VEB is employed and that we disregard OAM exchange.

We find that the line current density sources mimicking the well-focused excitation by a VEB centered at $\mathbf{R}_\mathrm{c}$ are expressed as
\begin{align}
\mathbf{J}_{\lbrace \mathrm{e/m}\rbrace}=J_{\lbrace \mathrm{e/m}\rbrace}\,\mathrm{e}^{\frac{\mathrm{i}\omega z}{v}}\delta(\mathbf{R}-\mathbf{R}_\mathrm{c})\mathbf{e}_z,
\label{Eq:Je}
\end{align} 
where $J_\mathrm{e}=-e$ is the amplitude of the electric current density and $J_\mathrm{m}$ is a (complex) amplitude of the effective magnetic current density to be determined. By inserting the current densities from Eq.~\eqref{Eq:Je} into Eqs.~\eqref{eq:del} and \eqref{eq:dmag}, 
we can write down an analogue of Eq.~\eqref{Eq:Eels_sum}, expressing the total loss probability of the sum of the four contributions
\begin{equation}
    \Gamma=\Gamma_\mathrm{e,\mathbf{J}_\mathrm{e}}+\Gamma_\mathrm{e,\mathbf{J}_\mathrm{m}}+\Gamma_\mathrm{m,\mathbf{J}_\mathrm{m}}+\Gamma_\mathrm{m,\mathbf{J}_\mathrm{e}},\label{eq:total_classical}
\end{equation}
where
\begin{equation}
    \Gamma_\mathrm{e,\mathbf{J}_\mathrm{\lbrace e/m\rbrace}}(\omega)=\frac{e}{\pi\hbar\omega}\int\limits_{-\infty}^\infty\mathrm{d}z\,\mathrm{Re}\left[E_{z,\mathbf{J}_\mathrm{\lbrace e/m\rbrace}}^\mathrm{ind}(\mathbf{R}_\mathrm{c},z,\omega)\, \mathrm{e}^{-\frac{\mathrm{i}\omega z}{v}}\right],\label{Eq:eEELS}
\end{equation}
\begin{equation}
    \Gamma_\mathrm{m,\mathbf{J}_\mathrm{\lbrace e/m\rbrace}}=\frac{-1}{\pi \hbar \omega}\int\limits_{-\infty}^\infty\mathrm{d}z\,\mathrm{Re}\left[B_{z,\mathbf{J}_\mathrm{\lbrace e/m\rbrace}}^\mathrm{ind}(\mathbf{R}_\mathrm{c},z,\omega)\,J^\ast_\mathrm{m}\mathrm{e}^{-\frac{\mathrm{i}\omega z}{v}}\right]\label{Eq:mEELS}.
\end{equation}

Here we split the induced electric field excited by the electric and the magnetic current ($\mathbf{E}_{\mathbf{J}_\mathrm{e}}^\mathrm{ind}$ and $\mathbf{E}_{\mathbf{J}_\mathrm{m}}^\mathrm{ind}$), yielding the corresponding loss probabilities $\Gamma_\mathrm{e,\mathbf{J}_\mathrm{e}}$ and $\Gamma_\mathrm{e,\mathbf{J}_\mathrm{m}}$, respectively. Similarly, the induced magnetic field originates from the interaction of the sample with both current sources  ($\mathbf{B}_{\mathbf{J}_\mathrm{e}}^\mathrm{ind}$ and $\mathbf{B}_{\mathbf{J}_\mathrm{m}}^\mathrm{ind}$), giving rise to the loss channels $\Gamma_\mathrm{m,\mathbf{J}_\mathrm{e}}$ and $\Gamma_\mathrm{m,\mathbf{J}_\mathrm{m}}$. This is denoted schematically in Fig.~\ref{fig:FigS2}(a).

\subsubsection*{Loss probability for a VEB interacting with a point-like dipolar particle in a semi-classical model}
\label{Sec:dipole_classical}

\begin{figure}
    \centering
    \includegraphics[width=\linewidth]{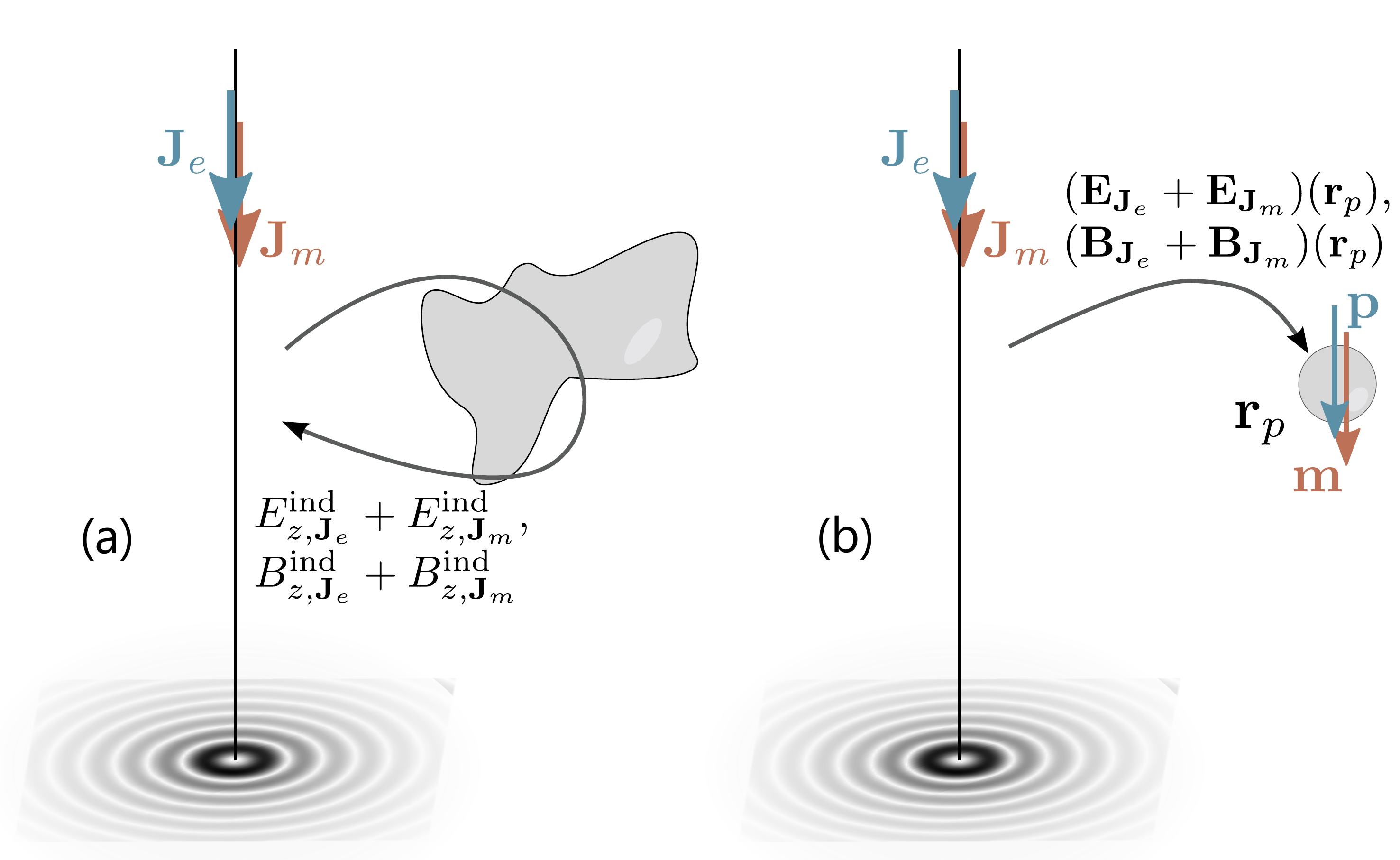}
    \caption{Illustration of the classical frameworks for calculating the interaction of the VEB with an environment, in which the electron beam is modelled as classical electric and magnetic, axial current densities $\mathbf{J}_\mathrm{e}$ and $\mathbf{J}_\mathrm{m}$. In (a) the electric fields generated by the currents and scattered by the environment are acting back on the currents, inducing loss [see Eqs.~\eqref{eq:total_classical}]. In (b) we consider an example of such interaction, with the environment modelled as a dipolar point-like scatterer, as in Fig.~\ref{fig:FigS1}. The energy loss experienced by the VEB is calculated from the interaction between the dipoles $\mathbf{p}$ and $\mathbf{m}$ induced in the scatterer, and the fields which induce the polarizations, generated by the VEB currents (see Section~\ref{Sec:dipole_classical}).}
    \label{fig:FigS2}
\end{figure}

To identify the correct expression for the magnetic current $\mathbf{J}_\mathrm{m}$ amplitude, we now aim to find the correspondence between the total loss given in Eq.~\eqref{eq:total_classical}, and the result considering the quantum-mechanical description of the VEB given in Eq.~\eqref{Eq:Gamapprox}. To this end, we consider here the same problem of scattering on a point-like dipolar particle as discussed in Section~\ref{Sec:dipole}.

The calculations are carried out in a different manner than above, shown schematically in Fig.~\ref{fig:FigS2}(b). Here we consider the electric and magnetic dipolar moments, $\mathbf{p}$ and $\mathbf{m}$ respectively, induced \textit{in the scatterer} $\mathbf{p}=\hat{\alpha}_\mathrm{EE}(\mathbf{E}_{\mathbf{J}_\mathrm{e}}+\mathbf{E}_{\mathbf{J}_\mathrm{m}})+\hat{\alpha}_\mathrm{EM}(\mathbf{B}_{\mathbf{J}_\mathrm{e}}+\mathbf{B}_{\mathbf{J}_\mathrm{m}})$ and $\mathbf{m}=\hat{\alpha}_\mathrm{MM}(\mathbf{B}_{\mathbf{J}_\mathrm{e}}+\mathbf{B}_{\mathbf{J}_\mathrm{m}})+\hat{\alpha}_\mathrm{ME}(\mathbf{E}_{\mathbf{J}_\mathrm{e}}+\mathbf{E}_{\mathbf{J}_\mathrm{m}})$, by the fields generated by the electric and magnetic currents of the VEB:
\begin{equation}
    \mathbf{E}(\mathbf{r}_\mathrm{p}) = \mathbf{E}_{\mathbf{J}_\mathrm{e}}(\mathbf{r}_\mathrm{p}) + \mathbf{E}_{\mathbf{J}_\mathrm{m}}(\mathbf{r}_\mathrm{p}),
\end{equation} 
\begin{equation}
    \mathbf{B}(\mathbf{r}_\mathrm{p}) = \mathbf{B}_{\mathbf{J}_\mathrm{e}}(\mathbf{r}_\mathrm{p}) + \mathbf{B}_{\mathbf{J}_\mathrm{m}}(\mathbf{r}_\mathrm{p}),
\end{equation}
where
\begin{subequations}
    \begin{align}
    &\mathbf{E}_{\mathbf{J}_\mathrm{e}}(\mathbf{r},\omega)\nonumber\\
    &=\frac{e\omega\mathrm{e}^{\frac{\mathrm{i}\omega z}{v}}}{2\pi\varepsilon_0\gamma v^2}\left[-K_1\left(\frac{\omega R}{v\gamma}\right)\mathbf{e}_R+\frac{\mathrm{i}}{\gamma}K_0\left(\frac{\omega R}{v\gamma}\right)\mathbf{e}_z\right],
    \end{align}
    \begin{align}
    \mathbf{B}_{\mathbf{J}_\mathrm{e}}(\mathbf{r},\omega)=\frac{-e\omega\mathrm{e}^{\frac{\mathrm{i}\omega z}{v}}}{2\pi\varepsilon_0\gamma v c^2}K_1\left(\frac{\omega R}{v\gamma}\right)\mathbf{e}_\phi,
    \end{align}
    \label{Eq:JE_Field}
\end{subequations}
and
    \begin{subequations}
    \begin{align}
    &\mathbf{B}_{\mathbf{J}_\mathrm{m}}(\mathbf{r},\omega)\nonumber\\
    &=\frac{-J_\mathrm{m}\omega\mathrm{e}^{\frac{\mathrm{i}\omega  z}{v}}}{2\pi\varepsilon_0\gamma v^2c^2}\left[-K_1\left(\frac{\omega R}{v\gamma}\right)\mathbf{e}_R+\frac{\mathrm{i}}{\gamma}K_0\left(\frac{\omega R}{v\gamma}\right)\mathbf{e}_z\right],
    \end{align}
    \begin{align}
    &\mathbf{E}_{\mathbf{J}_\mathrm{m}}(\mathbf{r},\omega)=\frac{-J_\mathrm{m}\omega\mathrm{e}^{\frac{\mathrm{i}\omega z}{v}}}{2\pi\varepsilon_0\gamma v c^2}K_1\left(\frac{\omega R}{v\gamma}\right)\mathbf{e}_\phi.
    \end{align}
    \label{Eq:JM_Field}
\end{subequations}

The loss of energy can be then calculated by considering the work done in the scatterers as:
\begin{align}
    \Gamma(\omega)=\frac{1}{\pi\hbar}\mathrm{Im}&\left[\mathbf{p}\cdot(\mathbf{E}^\ast_{\mathbf{J}_\mathrm{e}}(\mathbf{r}_\mathrm{p},\omega)+\mathbf{E}^\ast_{\mathbf{J}_\mathrm{m}}(\mathbf{r}_\mathrm{p},\omega))\right.\nonumber\\
    &\left.+\mathbf{m}\cdot(\mathbf{B}^\ast_{\mathbf{J}_\mathrm{e}}(\mathbf{r}_\mathrm{p},\omega)+\mathbf{B}^\ast_{\mathbf{J}_\mathrm{m}}(\mathbf{r}_\mathrm{p},\omega))\right].\label{Eq:Gam_dichr}
\end{align}

Plugging in Eqs.~\eqref{Eq:JE_Field} and \eqref{Eq:JM_Field} into the above expression, and axial polarizabilities as considered in Sec.~\ref{Sec:dipole}, we can find a closed expression for the loss probability: 
\begin{align}
    \Gamma&=\frac{\omega^2}{4\pi^3\varepsilon_0^2\hbar v^4\gamma^4}K_0^2\left(\frac{\omega R_\mathrm{p}}{v\gamma}\right)\mathrm{Im}\left[e^2\alpha_\mathrm{EE}+\alpha_\mathrm{MM} \frac{\left\lvert J_\mathrm{m}\right\rvert^2}{c^4}\right.\nonumber\\
    &\left.-e\alpha_\mathrm{EM}\frac{J_\mathrm{m}}{c^2}-e\alpha_\mathrm{ME}\frac{J^\ast_\mathrm{m}}{c^2}\right] \label{Eq:Loss_dipole_classical}.
\end{align}

If we compare Eq.~\eqref{Eq:Loss_dipole_classical} with Eq.~\eqref{Eq:Gamma_point_dipole_approx}, we find 
\begin{align}
    J_\mathrm{m}&=\frac{\mathrm{i}l\omega \,\mu_\mathrm{B}}{v},
    \label{Eq:Jm}
\end{align}
where we introduced Bohr magneton, $\mu_\mathrm{B}=e\hbar/(2 m_\mathrm{e})$ and set $l_\mathrm{i}=l$ as we are anyways disregarding the OAM exchange during the interaction.

In connection with our definition of different contributions to the loss probability, we can now see that the dichroic contribution stems from the \textit{crossed} interaction between the electrically induced magnetic response and the magnetic current and vice versa, contained in the terms $\Gamma_{\mathrm{m},\mathbf{J}_\mathrm{e}}$ and $\Gamma_{\mathrm{e},\mathbf{J}_\mathrm{m}}$, respectively.

\section{Loss probability for VEBs interacting with dielectric particles in the semi-classical approximation}
In the following, we present calculations of the loss probabilities [Eqs.~\eqref{Eq:eEELS} and \eqref{Eq:mEELS}] for different sample geometries where we solve for the induced EM field either analytically or numerically (as described in Appendix~\ref{AppC}). 

\subsection{Spectroscopy of localized modes in spherical dielectric nanoantennas}
We first apply the theory presented above to the canonical example of a single spherical nanoparticle. Due to its symmetry, the terms $\Gamma_{\mathrm{e},\mathbf{J}_\mathrm{m}}$ and $\Gamma_{\mathrm{m},\mathbf{J}_\mathrm{e}}$ do not contribute to the loss probability and we need to evaluate only the terms $\Gamma_{\mathrm{e},\mathbf{J}_\mathrm{e}}$ and $\Gamma_{\mathrm{m},\mathbf{J}_\mathrm{m}}$. The fully-retarded analytical solution of the induced electric field arising from the excitation of a spherical particle by an electric current was obtained in Ref.~\cite{garcia_sphere}, and the corresponding EEL probability is expressed as:

\begin{align}
\Gamma_{\mathrm{e}}^\mathrm{sph}\rightarrow\Gamma_{\mathrm{e,\mathbf{J}_\mathrm{e}}}&=\frac{e^{2}}{4\pi\varepsilon_0 c\hbar\omega}\sum\limits_{n=1}^{\infty}\sum\limits_{m=-n}^{n}K_{m}^{2}\left(\frac{\omega b}{v\gamma}\right)\nonumber\\
&\times\left[C_{n,m}^{\mathrm{M}}\mathrm{Im}[t_{n}^{\mathrm{M}}]+C_{n,m}^{\mathrm{E}}\mathrm{Im}[t_{n}^{\mathrm{E}}]\right],
\label{Eq:probability_sphere}
\end{align}
where the summation is performed over multipoles $(n,m)$, $K_m(x)$ is the modified Bessel function of the second kind of order $m$, $b$ is the distance of the beam from the center of the sphere (the impact parameter), and the coefficients $C_{n,m}^\mathrm{E/M}$ take into account the coupling with the field of the electron beam (see Eqs.~(30) and (31) of Ref.~\cite{garcia_sphere}). We use superscripts M/E to denote the coefficients related to the excitation of the magnetic/electric modes. Eq.~\eqref{Eq:probability_sphere} also includes the Mie coefficients:

\begin{align}
\label{tecka1}
t_{n}^{\mathrm{M}}&=\mathrm{i}\frac{j_{n}(k a)[k_{\mathrm{in}}a j_{n}(k_{\mathrm{in}}a)]'-j_{n}(k_{\mathrm{in}}a)[k a j_{n}(k a)]'}{h_{n}^{(1)}(k a)[k_{\mathrm{in}}a j_{n}(k_{\mathrm{in}}a)]'-j_{n}(k_{\mathrm{in}}a)[k a h_{n}^{(1)}(k a)]'},\\
\label{tecka2}
t_{n}^{\mathrm{E}}&=\mathrm{i}\frac{j_{n}(k a)[k_{\mathrm{in}}a j_{n}(k_{\mathrm{in}}a)]'-\varepsilon j_{n}(k_{\mathrm{in}}a)[k a j_{n}(k a)]'}{h_n^{(1)}(k a)[q_{\mathrm{in}}a j_n(k_{\mathrm{in}}a)]'-\varepsilon j_{n}(k_{\mathrm{in}}a)[k a h_n^{(1)}(k a)]'},
\end{align}
where $k_{\mathrm{in}}=\sqrt{\varepsilon}\omega/c$ the wave vector inside the sphere characterized by the relative dielectric function $\varepsilon$, and $a$ is the radius of the sphere. $j_{n}(x)$ and $h_{n}^{(1)}(x)$ are the spherical Bessel and Hankel functions of the first kind, respectively. The derivatives in Eq.~\eqref{tecka1} and Eq.~\eqref{tecka2} are performed with respect to the functions' arguments. 

When we consider the excitation of the sphere by a magnetic current, the corresponding loss probability $\Gamma_\mathrm{m}^\mathrm{sph}$ can be readily obtained by utilizing the transformation given by Eq.~\eqref{Eq:invariance_Maxwell}. The magnetic-current-mediated loss probability is thus given by

\begin{align}
&\Gamma_\mathrm{m}^\mathrm{sph}\rightarrow\Gamma_{\mathrm{m},\mathbf{J}_\mathrm{m}}=\left(\frac{l\,\mu_\mathrm{B}}{v}\right)^2\frac{\mu_0\omega}{4\pi c \hbar}\nonumber\\ &\times\sum\limits_{n=1}^{\infty}\sum\limits_{m=-n}^{n}K_{m}^{2}\left(\frac{\omega b}{v\gamma}\right)
\left[C_{n,m}^{\mathrm{M}}\mathrm{Im}[t_{n}^{\mathrm{E}}]+C_{n,m}^{\mathrm{E}}\mathrm{Im}[t_{n}^{\mathrm{M}}]\right].
\label{Eq:mag_probability_sphere}
\end{align}

\begin{figure}
\centering
\includegraphics[width=\linewidth]{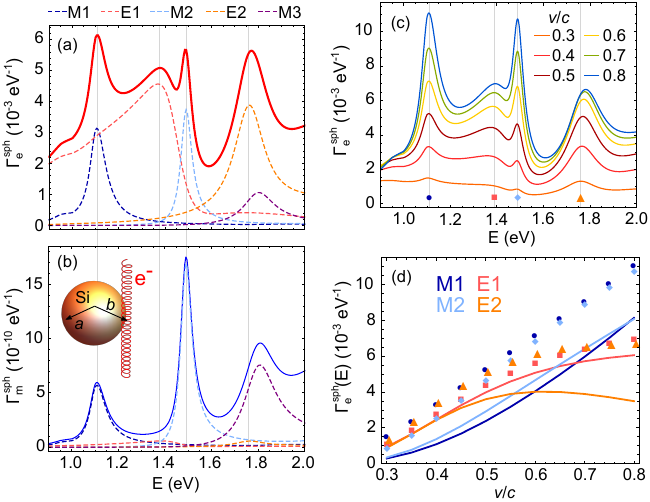}
\caption{(a) EEL probability, $\Gamma_\mathrm{e}^\mathrm{sph}$, as calculated by evaluation of Eq.~\eqref{Eq:probability_sphere} (solid red line) for an electron of energy 100~keV ($v=0.548 c$) interacting with a silicon spherical particle of radius $a=150$~nm at an impact parameter $b=157.5$~nm (see the inset of (b) for the geometrical arrangement). The dielectric function of amorphous silicon is taken from Ref.~\cite{palik}. (b) Magnetic-current mediated loss probability, $\Gamma_\mathrm{m}^\mathrm{sph}$, as calculated by evaluation of 
Eq.~\eqref{Eq:mag_probability_sphere}, 
for the same parameters 
as in (a), and considering OAM $\hbar l=100\hbar$ (solid blue line). Contributions to the loss probabilities from different electromagnetic modes are split in (a,b): magnetic dipole (M1, dark blue dashed line), electric dipole (E1, light red dashed line), magnetic quadrupole (M2, light blue dashed line), electric quadrupole (E2, orange dashed line), and magnetic octupole (M3, purple dashed line). Notice the different intensity scale of (b). (c) $\Gamma_\mathrm{e}^\mathrm{sph}$ evaluated for varying electron's velocity $v$. (d) Intensities of the peaks 
marked in (c) 
(colored symbols), with the color-coding 
corresponding to the dashed lines in (a,b). 
The contribution to the intensity
of each electromagnetic mode is also displayed by solid lines with the same color code as in (a,b).
\label{Fig1}}
\end{figure}

We now explore whether we can distinguish modes of electric and magnetic nature excited in silicon nanoparticles with the help of the v-EEL spectra. In Fig.~\ref{Fig1}(a,b) we show the calculated spectral contributions $\Gamma_\mathrm{e}^\mathrm{sph}$ [Eq.~\eqref{Eq:probability_sphere}; solid red line] and $\Gamma_\mathrm{m}^\mathrm{sph}$ [Eq.~\eqref{Eq:mag_probability_sphere}; solid blue line], respectively, for a single silicon nanosphere with radius $a=150$~nm, an impact parameter $b=157.5$~nm [as depicted in the inset of Fig.~\ref{Fig1}(b)] and a 100~keV beam ($v=0.548 c$). We note that while the spectrum in (a) does not depend on $\hbar l$ and thus is identical for a vortex and non-vortex beam, $\Gamma_\mathrm{m}$ in (b) is nonzero only for a vortex beam. We considered $\lvert l\rvert=100$ for our vortex beam.

To understand the origin of the resulting spectral features, we split the full spectra (solid lines) into the contributions of the different  electric (E$_n$) or magnetic (M$_n$) $n$-order multipoles (dashed lines), \textit{i.e.}, the spectra in Eqs.~\eqref{Eq:probability_sphere} and \eqref{Eq:mag_probability_sphere} before the summation over $n$. In the considered spectral range, the probability $\Gamma_\mathrm{e}^\mathrm{sph}$ [solid red line in Fig.~\ref{Fig1}(a)] exhibits four well-distinguishable peaks, which arise due to the excitation of a magnetic dipolar mode (M1, dark blue dashed line), electric dipole (E1, light red dashed line), magnetic quadrupole (M2, light blue dashed line) and electric quadrupole (E2, orange dashed line), whose energy nearly coincides with the magnetic octupole (M3, purple dashed line). On the other hand, due to the interchange of the coupling coefficients [compare Eq.~\eqref{Eq:probability_sphere} \textit{vs}. Eq.~\eqref{Eq:mag_probability_sphere}], only the magnetic modes (M1, M2 and M3) are found in the plot of $\Gamma_\mathrm{m}^\mathrm{sph}$ [solid blue line in Fig.~\ref{Fig1}(b)]. The cross-coupling of the magnetic current to the electric modes is negligible and produces only a small contribution [see dashed light red and orange lines in Fig.~\ref{Fig1}(b) close to zero]. 

In a typical measurement of EELS, 
one obtains the total loss probability $\Gamma^\mathrm{sph}=\Gamma_\mathrm{e}^\mathrm{sph}+\Gamma_\mathrm{m}^\mathrm{sph}$ [sum of solid spectra in (a) and (b)]. 
Therefore, in
order to separate the loss probability components $\Gamma_\mathrm{e}^\mathrm{sph}$ and $\Gamma_\mathrm{m}^\mathrm{sph}$, two measurements would be 
needed: one with a beam where $l\neq 0$ and another one with exactly the same experimental conditions
with a non-vortex beam ($l=0$). After subtracting these two spectra, one would obtain $\Gamma_\mathrm{m}^\mathrm{sph}$, which only shows 
the peaks corresponding to the magnetic modes. Unfortunately, we can observe that even for relatively large OAM, the \textit{magnetic} part of the loss probability $\Gamma_\mathrm{m}^\mathrm{sph}$ is six orders of magnitude smaller than $\Gamma_\mathrm{e}^\mathrm{sph}$, and thus falls below the limit of the currently achievable signal-to-noise ratio in STEM-EELS experiments. 

Besides varying the OAM of the VEB, there is another degree of freedom, which might be used to assign the spectral peaks to the modes as either electric or magnetic: the electron's speed $v$, which governs the strength of the coupling coefficients $C_{n,m}^\mathrm{E/M}$ related to the electromagnetic field of the fast electrons. In Fig.~\ref{Fig1}(c) we evaluate $\Gamma_\mathrm{e}^\mathrm{sph}$ for varying $v$ and $l=0$ (conventional electron beam). We observe that the intensity ratio of the four visible peaks changes significantly. With increasing accelerating voltage (electron's speed), the coupling of the beam with the magnetic modes is much more efficient, which results from the fact that the accompanying magnetic field is stronger for faster electrons. Further, the 
intensity corresponding to the excitation of the M1 and M2 modes grows faster than the peak assigned to the E1 mode, which starts to saturate for larger speeds ($v>0.7 c$). This trend is confirmed in Fig.~\ref{Fig1}(d), where we plot the intensities of the peaks extracted from spectra in Fig.~\ref{Fig1}(c) at the energies corresponding to the M1 (dark blue points), E1 (light red squares), M2 (light blue diamonds), and E2 (orange triangles) modes, depending on the electron's speed. We also plot the peak intensities as if the modes were excited independently by solid lines to eliminate the influence of the spectral overlap of the excited modes [see Fig.~\ref{Fig1}(a) showing that, \textit{e.g.}, E1 contributes significantly even at the energy of the M1 peak]. This trend is similar for higher-order modes, and we suggest that obtaining the EEL spectra at several acceleration voltages might serve for a relatively straightforward classification of the modes.

\begin{figure}
\centering
\includegraphics[width=\linewidth]{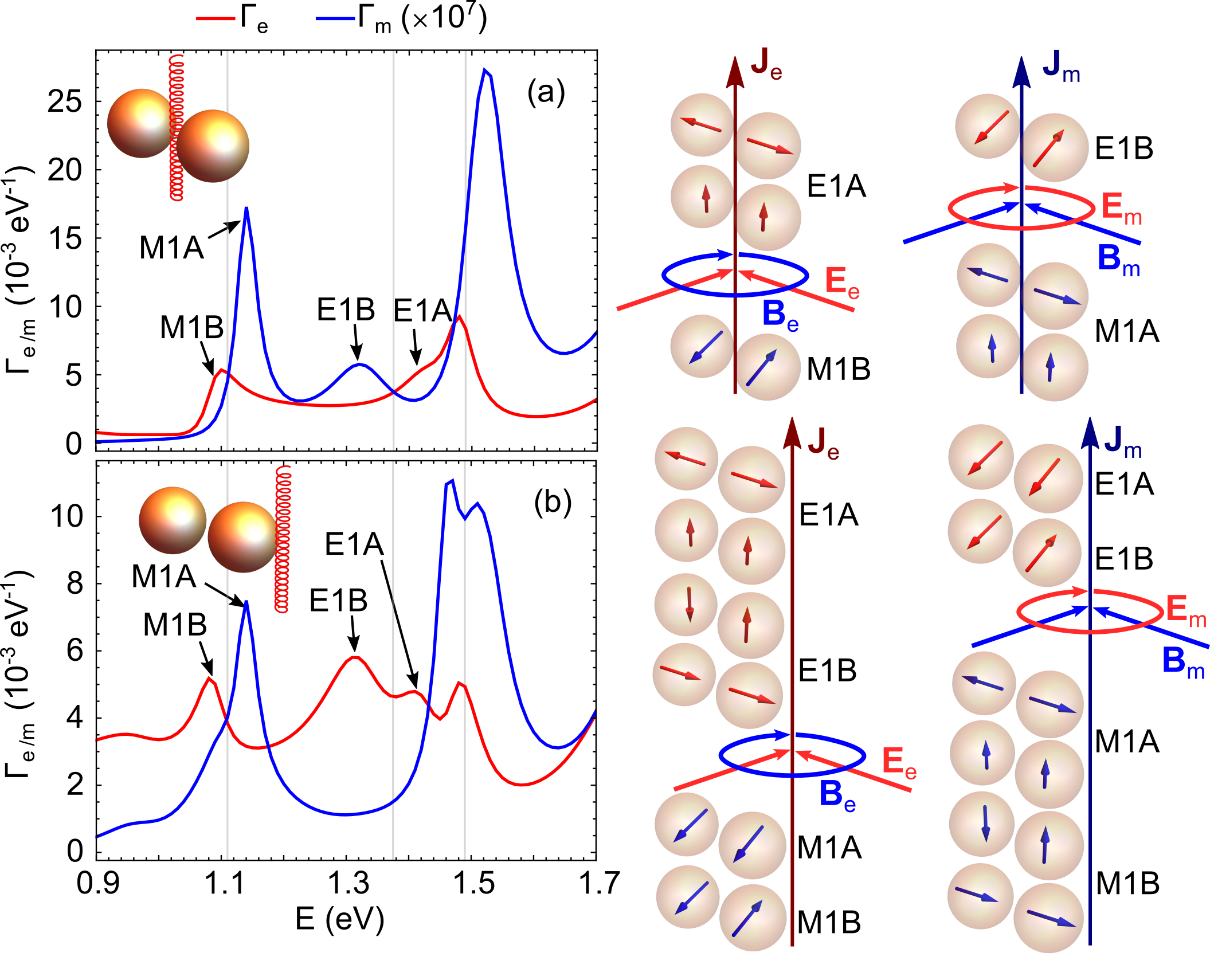}
\caption{(a,b) Numerically calculated electric-current and magnetic-current-mediated EEL probabilities, $\Gamma_\mathrm{e}$ (solid red line) from Eq.~\eqref{Eq:eEELS}, and $\Gamma_\mathrm{m}$ (solid blue line) from Eq.~\eqref{Eq:mEELS}, respectively, for a 100-keV electron beam with $l=100$ exciting a spherical particle dimer. Each spherical particle of radius $a=150$~nm is made of silicon (dielectric response is taken from Ref.~\cite{palik}). The distance of the gap between the particles is $g=15$~nm and the beam is passing either (a) through the middle of the gap or (b) by the side of one of the particles at the axis of the dimer, 7.5~nm from its surface. We denote the loss peaks corresponding to the hybridized dipolar modes in the dimer as bonding and antibonding magnetic dipole (M1B and M1A), and bonding and antibonding electric dipole (E1B and E1A). Gray vertical lines denote spectral positions of the modes M1, E1 and M2 in the EEL spectra calculated for a single spherical particle (see Fig.~\ref{Fig1}). Next to the corresponding spectra, electric (red arrows) and magnetic (blue arrows) coupled-dipole configurations excitable by each of the current contributions is schematically depicted.
\label{Fig2}}
\end{figure}

VEBs can also be applied to unravel the spectral response of more complex dielectric nanostructures, such as dimers of two (identical) particles separated by a small gap. The nanoparticle dimers are also of large interest as they can provide a significant enhancement of the field in the gap or yield directional scattering \cite{fu_Si_directional,albella_Surface-enahnced,bakker,Yan_Fano}. In Fig.~\ref{Fig2} we thus study numerically v-EELS of a pair of 
spherical dielectric particles (each of them with the same properties as 
the single spherical particle studied in Fig.~\ref{Fig1}) separated by a gap 
of distance 
$g=15$~nm. It has been shown that in such a system, the modes of the individual particles hybridize and form bonding and antibonding modes of the dimer \cite{Zywietz}. In Fig.~\ref{Fig2}(a,b) we analyze how these 
hybridized modes contribute to the spectra for different VEB positions. 

In Fig.~\ref{Fig2}(a,b) we consider a 100-keV electron beam with $\lvert l\rvert=100$ passing through the middle of the gap or close to the side of one of the spheres and calculate the EEL probability. We note that the \textit{crossed} loss probability components $\Gamma_{\mathrm{e},\mathbf{J}_\mathrm{m}}$ and $\Gamma_{\mathrm{m},\mathbf{J}_\mathrm{e}}$ are either identically zero or cancel. We can thus again assign $\Gamma_\mathrm{e}\rightarrow\Gamma_{\mathrm{e},\mathbf{J}_\mathrm{e}}$ and $\Gamma_\mathrm{m}\rightarrow\Gamma_{\mathrm{m},\mathbf{J}_\mathrm{m}}$. Importantly, the 
symmetry of the electric or magnetic field produced by the electric or magnetic part of VEB current dictates which current component couples to specific modes of the dimer. We schematically depict possible scenarios next to the graph. 

If the electron beam is placed in the gap [Fig.~\ref{Fig2}(a)], $\Gamma_\mathrm{e}$ [solid red line in Fig.~\ref{Fig2}(a)] shows that the electric current component can excite the magnetic dipolar bonding mode (M1B), the electric antibonding mode (E1A), and the bonding magnetic quadrupolar mode yielding a peak close to 1.5~eV. On the other hand, the magnetic part of the current couples to the magnetic dipolar anti-bonding mode (M1A), the electric dipolar bonding mode (E1B) and the anti-bonding magnetic quadrupolar mode (see the peak above 1.5~eV), which appears in $\Gamma_\mathrm{m}$ [solid blue line in Fig.~\ref{Fig2}(a)]. The energy splitting of the bonding and anti-bonding modes is apparent when the peak positions are compared to the spectral positions of the modes excited in the individual sphere (plotted by vertical gray lines, extracted from 
Fig.~\ref{Fig1}). 

When the beam is moved to the side of one of the spheres along the dimer axis [see the schematics in Fig.~\ref{Fig2}(b)], bonding and anti-bonding dipolar modes are excitable by both current components as schematically 
shown next to the graph. However, some of the dipolar arrangements are excited preferentially, which is apparent 
in the respective spectra. The electric component of the current efficiently couples with the M1B mode, E1A mode, and also E1B mode [solid red line in Fig.~\ref{Fig2}(b)]. On the other hand, $\Gamma_\mathrm{m}$ [solid blue line in Fig.~\ref{Fig2}(b)] shows a spectral feature arising from the excitation of both M1A and M1B. Interestingly, M1A is dominant with respect to M1B, whose excitation gives rise to a small shoulder below 1.1~eV. We also observe that the magnetic current component couples only weakly to the E1B and E1A modes. However, we note that this loss contribution is still 
six to seven orders of magnitude smaller than the electric part and would be difficult to isolate, as discussed above.

\subsection{Probing the photonic density of states in an infinite cylinder}
\label{Sec:cylinder}
\begin{figure*}
	\centering
	\includegraphics[width=0.7\linewidth]{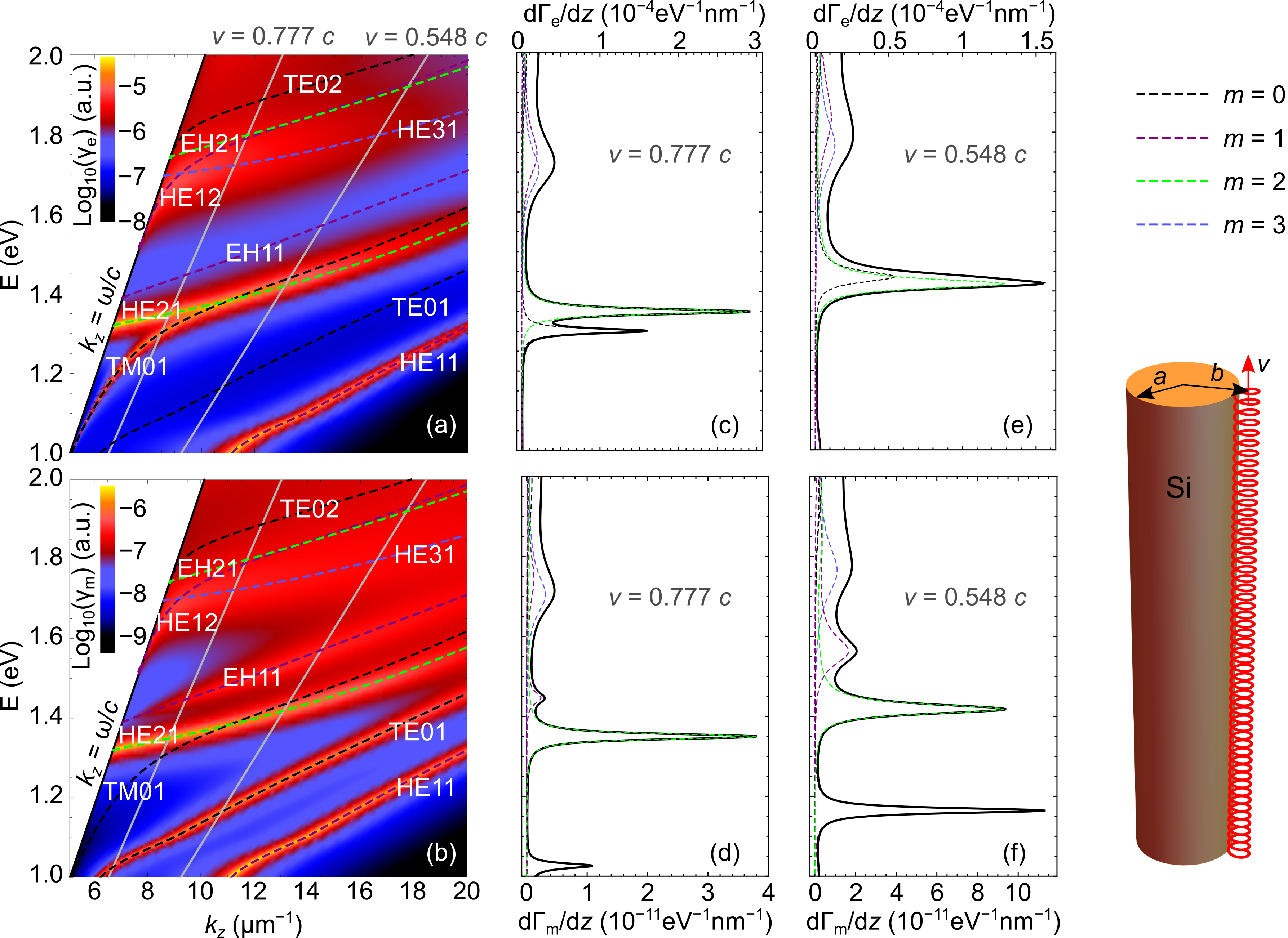}
	\caption{EEL probability calculations for an electron beam moving parallel to the axis of an infinite silicon cylinder along the $z$ direction (see the schematics) with speed $v$. The radius of the cylinder is $a=150$~nm and the beam distance from the center of the cylinder is $b=160$~nm. (a,b) Electric- and magnetic-current mediated $(k_z,\omega)$-dependent loss probabilities $\gamma_\mathrm{e}$ from Eq.~\eqref{Eq:cylinder_gam_el} and $\gamma_\mathrm{m}$ from \eqref{Eq:cylinder_gam_mag} in logarithmic color scale. Dashed lines overlaying the density plot correspond to dispersions of different azimuthal modes $m$ from Eq.~\eqref{Eq:cylinder_dispersion} (color coding shown in the legend). Solid gray lines show the wavevectors provided by the electron beam, $k_z=\omega/v$ for velocities $v=0.777c$ (300-keV beam) and $v=0.548c$ (100-keV beam). (c,d) Loss probabilities per unit trajectory corresponding to the electric current component [Eq.~\eqref{Eq:cylinder_gam_el}] and to the magnetic current component [Eq.~\eqref{Eq:cylinder_gam_mag}] calculated for a 300-keV beam (solid black line) and $l=100$. The dashed lines show contributions of different azimuthal components $m$ to the spectra. (e,f) the same as (c,d) but for a 100-keV beam. For plots (a,b) we considered only the first four azimuthal modes in the summation. The total probabilities in (c-f) were calculated with $m=0,1,...,6$.
		\label{VEB_Fig3}}
\end{figure*}

Another canonical example of a dielectric system with a strong electric and magnetic response, which can be characterized via v-EELS, is that of dielectric waveguides \cite{rybin2015switching,holsteen2017purcell,cihan2018silicon,traviss2015antenna,abujetas,cao2009engineering}. Previous theoretical analysis of the interaction of fast electrons with dielectric cylindrical waveguides has already suggested their potential for 
applications in single-photon sources \cite{bendana}. Here we study the EEL probability of a VEB exciting an infinite cylindrical wire of radius $a$ placed in vacuum for a geometrical arrangement as sketched in Fig.~\ref{VEB_Fig3}: an electron beam moving at speed $v$ parallel to the axis of the wire at a 
distance $b>a$ from the center of the cylinder. For this geometry, the retarded analytical solution of the EEL probability was presented \textit{e.g.}~in Ref.~\cite{walsh} and reproduced in Appendix~\ref{App:cylinder}, which we can easily modify to include 
the contribution to the loss experienced by the magnetic component of the current by 
using the transformation in Eq.~\eqref{Eq:invariance_Maxwell}. We can write the two contributions to 
the overall loss probability of the VEB per unit length as:

\begin{widetext}
\begin{align}
\frac{\mathrm{d}\Gamma_\mathrm{e}}{\mathrm{d}z}&=\frac{\mathrm{d}\Gamma_{\mathrm{e},\mathbf{J}_{e}}}{\mathrm{d}z}=\frac{e^2}{2\pi^2\hslash\omega^2\varepsilon_0}\underbrace{\sum_{m=0}^\infty\mathrm{Re}\left\lbrace(2-\delta_{m0})K_m\left(\kappa_\mathrm{o} b\right)(k_z^2-k^2)b_{\mathrm{e},m}(k_z,\omega)\right\rbrace}_{\gamma_\mathrm{e}(k_z,\omega)}\rvert_{k_z=\frac{\omega}{v}}, \label{Eq:cylinder_gam_el}\\
\frac{\mathrm{d}\Gamma_\mathrm{m}}{\mathrm{d}z}&=\frac{\mathrm{d}\Gamma_{\mathrm{m},\mathbf{J}_{m}}}{\mathrm{d}z}=\frac{\mu_\mathrm{B}^2 l^2\mu_0}{2\pi^2 v^2\hbar}\underbrace{\sum_{m=0}^\infty\mathrm{Im}\left\lbrace(2-\delta_{m0})K_m\left(\kappa_\mathrm{o}b\right)(k_z^2-k^2)d_{\mathrm{m},m}(k_z,\omega)\right\rbrace}_{\gamma_\mathrm{m}(k_z,\omega)}\rvert_{k_z=\frac{\omega}{v}} \label{Eq:cylinder_gam_mag},
\end{align}
\end{widetext}
where $m$ denotes different azimuthal modes, $\delta_{m0}$ is the Kronecker delta, and 
$k_z=\omega/v$ stands for the wavevector along the cylinder axis, which has to match the wavevector component transferred from the fast electron when calculating the spectra. The dimensionless coefficients $b_{\mathrm{e},m}(k_z,\omega)$ and $d_{\mathrm{m},m}(k_z,\omega)$ can be obtained as described in Appendix~\ref{App:cylinder}. We also define  $\kappa_\mathrm{o}=\sqrt{k_z^2-k^2}$.

The denominators of the coefficients $b_{\mathrm{e},m}$ and $d_{\mathrm{m},m}$ yield the dispersion of all the modes supported by the cylinder with the relative dielectric function $\varepsilon$:
\begin{widetext}
\begin{align}
\frac{k^2 k_z^2 m^2}{a^2\kappa_\mathrm{o}^4\kappa_\mathrm{i}^2}K_m(\kappa_\mathrm{o}a)^2 I_m(\kappa_\mathrm{i}a)^2(\varepsilon-1)^2&+\left[\frac{\kappa_\mathrm{i}}{\kappa_\mathrm{o}}I_m(\kappa_\mathrm{i}a)K'_m(\kappa_\mathrm{o}a)-K_m(\kappa_\mathrm{o}a)I'_m(\kappa_\mathrm{i}a)\right]\nonumber\\
&\times\left[\varepsilon K_m(\kappa_\mathrm{o}a)I'_m(\kappa_\mathrm{i}a)-\frac{\kappa_\mathrm{i}}{\kappa_\mathrm{o}}I_m(\kappa_\mathrm{i}a)K'_m(\kappa_\mathrm{o}a)\right]=0,
\label{Eq:cylinder_dispersion}
\end{align}
\end{widetext}
where $I_m(x)$ and $K_m(x)$ are the modified Bessel functions of the first and the second kind, respectively, of order $m$, and $\kappa_\mathrm{i}=\sqrt{k_z^2-k_\mathrm{in}^2}$.

In Fig.~\ref{VEB_Fig3}(a,b) we plot the $(k_z,\omega)$-dependent loss probabilities $\gamma_\mathrm{e}$ and $\gamma_\mathrm{m}$ as implicitly defined in Eq.~\eqref{Eq:cylinder_gam_el} and Eq.~\eqref{Eq:cylinder_gam_mag}, respectively, for geometrical parameters $a=150$~nm and $b=160$~nm. On top of the density plots, we show the 
dispersion curves corresponding to the guided EM modes (the 
leaky modes above the light line, 
given by $k_z=\omega/c$, 
are not shown as they are not excitable by the parallel beam) supported by the infinite cylinder obtained as solutions of Eq.~\eqref{Eq:cylinder_dispersion} for the different orders $m$ denoting the modes' azimuthal symmetry. We consider only the first four azimuthal numbers $m=\lbrace 0,1,2,3\rbrace$, and 
as higher-order modes are much more damped, with these modes 
we can capture all the dominant spectral features. We can observe that some of the modes (denoted by using the standard notation from waveguide theory, see \textit{e.g.}~Refs.~\cite{Snitzer, Tong:04}) are visible only in the electrical contribution to the spectra, $\gamma_\mathrm{e}$ [Fig.~\ref{VEB_Fig3}(a)] or, \textit{vice versa}, in the magnetic contribution, $\gamma_\mathrm{m}$ [Fig.~\ref{VEB_Fig3}(b)]. We can conclude that due to the symmetry of the EM field produced by the current components, transverse electric modes (TE01, TE01,...) are excitable only by the magnetic current. On the contrary, transverse magnetic modes (TM01, TM02,...) couple only to the electric current component. 

From the $(k_z,\omega)$-dependent plots in Fig.~\ref{VEB_Fig3}(a,b) we can readily obtain the EEL spectra by setting $k_z=\omega/v$ [see the gray lines in Fig.~\ref{VEB_Fig3}(a,b)]. In Fig.~\ref{VEB_Fig3}(c,d) we plot $\mathrm{d}\Gamma_{\mathrm{e},\mathbf{J}_\mathrm{e}}/\mathrm{d}z$ [Eq.~\eqref{Eq:cylinder_gam_el}] and $\mathrm{d}\Gamma_{\mathrm{m},\mathbf{J}_\mathrm{m}}/\mathrm{d}z$ [Eq.~\eqref{Eq:cylinder_gam_mag}], respectively, evaluated for a 300-keV electron beam ($v=0.777 c$). The total probabilities (solid black lines) are split into contributions of the different azimuthal modes $m=\lbrace 0,1,2,3\rbrace$ denoted by the colored dashed lines. We observe that the hybrid HE21 mode produces the dominant spectral feature in both cases. On the other hand, the peak corresponding to the excitation of the TM01 mode is present only in the spectrum of Fig.~\ref{VEB_Fig3}(c), and the peak arising from the excitation of the TE01 mode appears only in Fig.~\ref{VEB_Fig3}(d). The hybrid EH11 is also dominantly excitable by the magnetic current and has only a negligible contribution in the electric-current-mediated spectrum, as confirmed by evaluating the induced fields given in Appendix~\ref{App:cylinder}.

By changing the acceleration voltage to 100~kV, we obtain the spectra in Fig.~\ref{VEB_Fig3}(e,f), where the same modes give rise to peaks at slightly different energies due to the change of the energy-momentum matching (higher $k_z$ is provided at fixed energy compared to the faster 300-keV electron). Importantly, the standard electrical component of the EEL probability in geometries possessing translational invariance, such as in the current situation of the beam moving parallel to an infinite cylinder, can be related to the electrical part of the projected photonic local density of states (LDOS)\cite{LDOS}. An analogous 
proportionality holds between the magnetic-current-mediated loss and the magnetic part of the photonic LDOS. 
Hence, this relationship might be used in the context of the interaction of magnetic emitters with such structures. Our results are consistent with the findings in Ref.~\cite{Verhart:14}, where the coupling of modes to differently oriented electric dipoles was linked to the excitation by electric/magnetic current components.

\subsection{Dichroic spectroscopy with vortex electron beams}

\begin{figure*}
	\centering
	\includegraphics[width=0.75\linewidth]{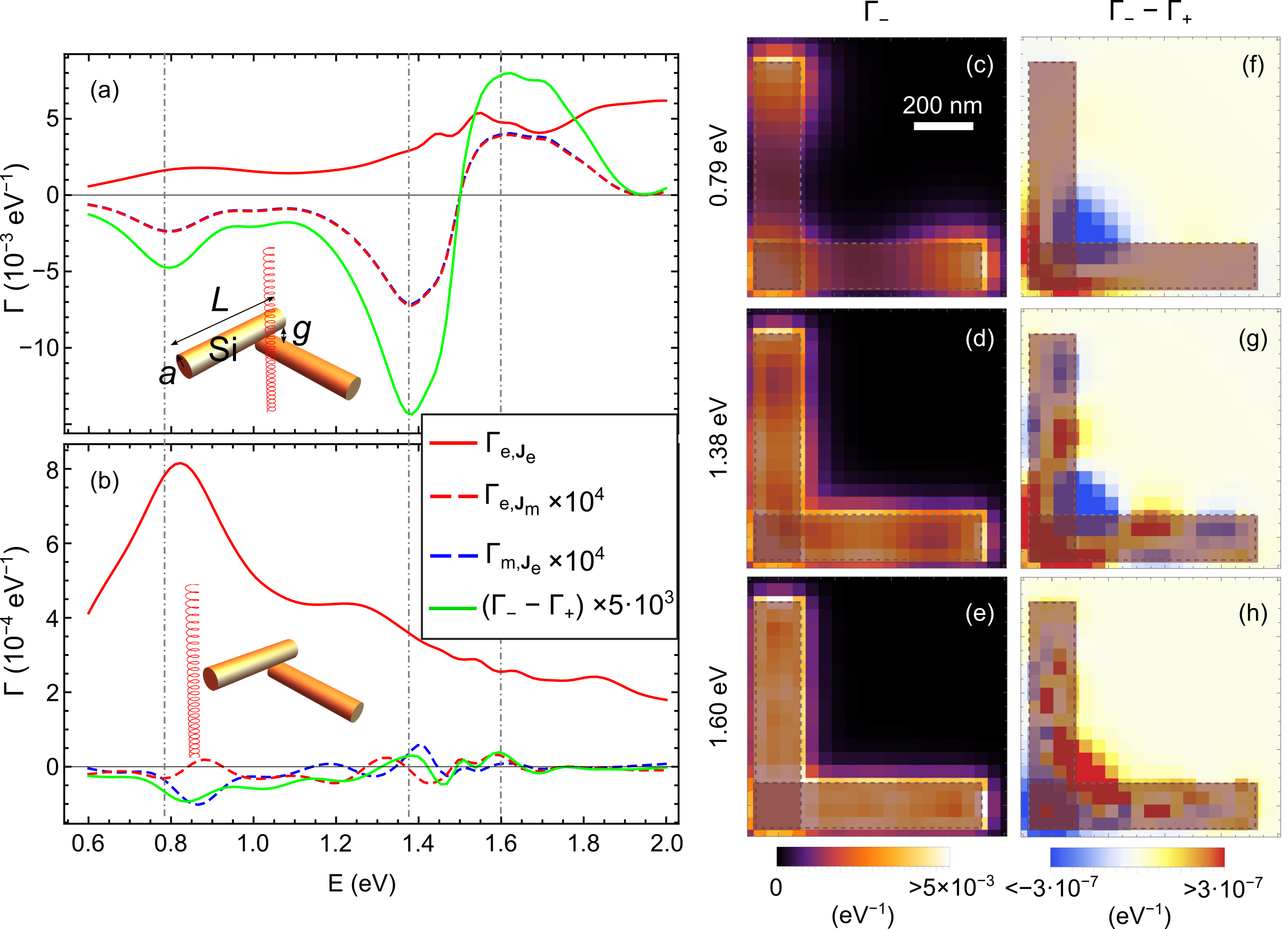}
	\caption{(a,b) v-EEL spectra calculated for chirally-arranged silicon rods with parameters $a=80$~nm, $L=800$~nm, $g=30$~nm [see the inset of (a)]. The dielectric response of silicon is taken from Ref.~\cite{palik}. We consider the excitation by a 100~keV VEB with $l=-100$ and split the spectral contributions according to Eqs.~\eqref{Eq:eEELS} and \eqref{Eq:mEELS}: $\Gamma_{\mathrm{e},\mathbf{J}_\mathrm{e}}$ (solid red line), $\Gamma_{\mathrm{e},\mathbf{J}_\mathrm{m}}$ (dashed red line), and $\Gamma_{\mathrm{m},\mathbf{J}_\mathrm{e}}$ (dashed blue line). We omit the $\Gamma_{\mathrm{m},\mathbf{J}_\mathrm{m}}$ term. The green line shows v-EEL dichroism obtained as a difference between spectra calculated for $l=-100$ and $l=+100$. In (a) the beam is positioned according to the inset 10~nm from the surface of the rods, whereas in (b) the beam is placed 10~nm from the tip of one of the rods as depicted. (c-e) Energy-filtered maps of the total EEL for $l=-100$, and (f-h) the dichroic signal at energies marked on the left of each row. The selected energies correspond to positions of the peaks in the dichroic signal in (a). The spatial scaling is the same in all maps and the boundaries of the cylinders are plotted by dashed gray lines while their cross sections are shown as semi-transparent rectangles.
		\label{Fig:VEB_chiral_Si}}
\end{figure*}

Now we demonstrate the emergence of the dichroic signal when an extended chirally-arranged nanostructure is probed by a VEB. We adopt a similar geometry as the one studied in Ref.~\cite{yin_chiral_rods}, and perform the numerical modeling for two vertically displaced cylindrical rods rotated by $90^{\circ}$ and stacked at their corners as shown in the 
schematics of Fig.~\ref{Fig:VEB_chiral_Si}. The 
overall response of the structure in such an arrangement yields optical dichroism \cite{yin_chiral_rods}, and thus, according 
to the preceding analysis, we should expect the emergence of dichroism also in the 
v-EEL spectra. However, the finite spatial extent of the structure, as well as the 
overlap of different electro-magnetic modes excited in the silicon rods, might produce a non-trivial spatial dependence of the dichroic signal.

We set the length of each rod, 
$L=800$~nm, radius $a=80$~nm and vertical spacing between the rods $g=30$~nm as shown in the inset of Fig.~\ref{Fig:VEB_chiral_Si}(a). We calculate EEL spectra for an excitation by a VEB with energy 100~keV and OAM $l=\pm 100$ at different beam positions. The spectra in Fig.~\ref{Fig:VEB_chiral_Si}(a) are obtained for the beam placed at the corner between the rods, 10~nm from their boundaries, whereas in  Fig.~\ref{Fig:VEB_chiral_Si}(b) the beam is located at 
10~nm from the tip of one of the rods (see the corresponding insets). We split the relevant spectral components as in Eqs.~\eqref{Eq:eEELS} and \eqref{Eq:mEELS}: the purely electric loss probability term $\Gamma_{\mathrm{e},\mathbf{J}_\mathrm{e}}$ (solid red line), and the \textit{crossed} electric-magnetic terms $\Gamma_{\mathrm{e},\mathbf{J}_\mathrm{m}}$ (dashed red line) and $\Gamma_{\mathrm{m},\mathbf{J}_\mathrm{e}}$ (dashed blue line). We do not plot the purely magnetic term $\Gamma_{\mathrm{m},\mathbf{J}_\mathrm{m}}$, which is several orders of magnitude weaker.

The modes of a finite silicon cylinder can be understood as standing waves along the long axis of the cylinder, formed by the different modes of an infinite cylinder \cite{traviss2015antenna,Ee_NL} discussed in Sec.~\ref{Sec:cylinder}. The first dominant peak in $\Gamma_{\mathrm{e},\mathbf{J}_\mathrm{e}}$ close to $0.8$~eV in Fig.~\ref{Fig:VEB_chiral_Si}(a,b) originates mostly from the bonding arrangement of two dipolar modes along both rods 
[as can be noted also in the energy-filtered map in Fig.~\ref{Fig:VEB_chiral_Si}(c)] with TM01 transverse modal profile. This mode couples well in this geometry to the EM field associated with 
the electric current component as found from the 
analysis of the induced field. The second-order mode with TM01 modal profile appears around 1.3~eV [the corresponding peak is clearly visible in Fig.~\ref{Fig:VEB_chiral_Si}(b)], but it significantly overlaps with an admixture of higher-order electric and magnetic modes from 
 higher energies. Hence, due to this overlap, the energy-filtered maps at energies 1.38~eV [Fig.~\ref{Fig:VEB_chiral_Si}(d)] and also at 
 1.6~eV [Fig.~\ref{Fig:VEB_chiral_Si}(e)] show nearly homogeneous intensity for all beam positions close to the rod surfaces.

Importantly, there is a small difference between the spectra $\Gamma_-$ and $\Gamma_+$, which we plot with solid green lines in Fig.~\ref{Fig:VEB_chiral_Si}(a,b). The dichroism in EEL emerges from the \textit{crossed} loss terms $\Gamma_{\mathrm{e},\mathbf{J}_\mathrm{m}}$ and $\Gamma_{\mathrm{m},\mathbf{J}_\mathrm{e}}$ [dashed lines in Fig.~\ref{Fig:VEB_chiral_Si}(a,b)]. As we can observe in the energy-filtered maps of $(\Gamma_- - \Gamma_+)$ [Fig.~\ref{Fig:VEB_chiral_Si}(f-h)], the strongest dichroic response arises for the beam close to the stacking point of the rods, where the strongest interference and the phase difference between the fields induced at each of the rods appear. Similar behavior was predicted in a recent work \cite{lourenco_NatPhys}. The dichroic v-EEL spectra also flip signs depending on the spatial distribution of the local phase and the nature of the induced field along the $z$ axis. The sign change of the dichroic signal appears in the region between 1.4~eV and 1.6~eV, where hybridized modes with TE polarization, \textit{i.e.} coupled magnetic dipoles and higher-order modes polarized along the long axes of the rods, can be excited. These modes couple preferentially to the EM field of the magnetic current component, which changes its sign depending on $l$ [see Eq.~\eqref{Eq:Jm}]. Hence, the sign of the dichroic signal might, in this case, reflect whether a particular mode preferentially couples with either the electric or the magnetic current component of the VEB. However, we note that interpreting spatially-resolved dichroic v-EELS in a general case can be rather involved and requires further theoretical analysis.

Although the dichroic signal is three to four orders of magnitude weaker than the overall spectra, experimental development and involvement of high OAM might make it detectable \cite{tavabi}. We note that our approach assuming an infinitely focused VEB presumably underestimates the intensity of the dichroic signal. By taking into account an overlap of a realistic beam profile with the electromagnetic field in the structure (\textit{e.g.}, as in Refs.~\cite{Asenjo_garcia,zanfrognini2018orbital}), one might expect a higher contribution of the dichroic signal to the overall loss probability, with a qualitatively similar spatial dependence. 

\section{Conclusions}
We set a classical theoretical framework suitable for qualitative modeling of vortex electron energy loss spectroscopy at optical frequencies, in the limit of a perfectly-focused vortex beam. We revealed that spatially-resolved EELS acquired with electron vortices could be a powerful technique for a detailed characterization of the optical response of complex nanostructures, which we demonstrated in several examples: spherical particles and cylindrical wires made of silicon. In particular, we showed how to interpret EEL spectra based on field symmetry considerations and demonstrated that we could distinguish modes of electric or magnetic nature emerging in the dielectric nanoparticles by varying electron's velocity or OAM. We also proved the emergence of dichroism in electron spectra recorded with vortex electrons, which could establish v-EELS as a unique technique to characterize chirality at the nanoscale. 

\begin{acknowledgments}
A.K. acknowledges the support of the Czech Science Foundation GACR under the Junior Star grant No. 23-05119M. R.H. was financially supported by the Spanish Ministry of Science and Innovation under the Mar\'{i}a de Maeztu Units of Excellence Program
(CEX2020-001038-M/MCIN/AEI/10.13039/501100011033) and the Project PID2021-123949OB-I00. J.A. acknowledges the Spanish Ministry of Economy, Industry and Competitiveness (project PID2019-107432GB-I00). M.K.S. acknowledges support from the Macquarie University Research Fellowship scheme (MQRF0001036) and the Australian Research Council Discovery Early Career Researcher Award DE220101272.
\end{acknowledgments}

\appendix
\section{VEB radius as a function of aperture size and OAM}\label{app:radius}
\begin{figure}[h!]
    \centering
    \includegraphics[width=\linewidth]{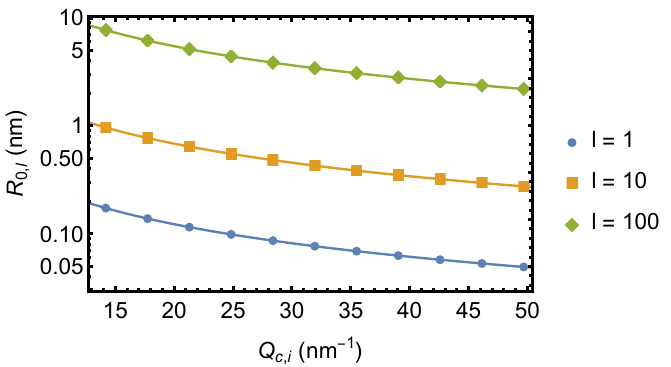}
    \caption{Effective beam radius as a function of of the initial transverse wavevector cutoff calculated for OAM of $\hbar$, $10\hbar$ and $100\hbar$. We find that numerically-calculated data (points) approximately follow $1/Q_\mathrm{c,i}$ dependence with the corresponding fits plotted as solid lines. In practice, there is a finite achievable radius given by the maximal possible convergence angle.}
    \label{FigS1}
\end{figure}

In Fig.~\ref{FigS1} we evaluate the dependence of the effective VEB radius for different forming aperture sizes and OAM.

\section{Note: Classical electric and magnetic current components}
\label{AppB} 
As we showed in the main text, we can find that the approximate electric current $\mathbf{J}_\mathrm{e}=-e \mathbf{W}(R)\mathrm{e}^{\mathrm{i}\omega z/v}$ can be equivalently replaced by two sources $\mathbf{J}_\mathrm{e}=-e\, \mathrm{e}^{\mathrm{i}\omega z/v}\mathbf{e}_z$ and 
\begin{align}
    \mathbf{J}_\mathrm{m}\approx\frac{\mathrm{i}le \omega}{2q_{z,\mathrm{i}}}\mathrm{e}^{\mathrm{i}\omega z/v}\mathbf{e}_z=\frac{\mathrm{i} l\omega\mu_\mathrm{B} }{v}\mathrm{e}^{\mathrm{i}\omega z/v}\mathbf{e}_z.
    \label{Eq:Jm_der}
\end{align}

However, we note that especially for larger $l$ or for probed structures with dimensions comparable to the vortex focus, the loss probability calculated using this model can differ both qualitatively and quantitatively from the rigorous approach, \textit{i.e.}~using the overlap integral in Eq.~\eqref{Eq:GamQ} within the quantum-mechanical description of the VEBs. We also note that an alternative magnetic current deduced from an effective spiralling electric current, as 
presented in Ref.~\cite{Mohammadi}, is expressed as 
$\mathbf{J}_\mathrm{m}=2\mathrm{i}l e\hbar c^2/(m_\mathrm{e}v R_{0,l}^2\omega)\mathrm{e}^{\mathrm{i}\omega z/v}\mathbf{e}_z$, which 
involves an additional factor $4c^2/(R_{0,l}^2\omega^2)$ in their expression, as 
compared to Eq.~\eqref{Eq:Jm_der}, yielding several orders of magnitude larger values for the magnetic field current amplitude. 

\section{Numerical calculations of (VEB-)EELS in Comsol Multiphysics}
\label{AppC}
We utilize the \texttt{Radio Frequency} toolbox of \texttt{Comsol Multiphysics} software where we solve the wave equation for the total electric and magnetic field in the frequency domain with electric and magnetic current sources. We perform the calculations in a 3D simulation domain in the Cartesian coordinate system $(x,y,z)$. The simulation domain includes the nanostructure characterized by a dielectric response $\varepsilon$, a straight line representing the electron's trajectory, a simulation domain (typically a block) surrounding the nanostructure characterized by $\varepsilon=1$ and \texttt{Perfectly Matched Layers} (PML) with Cartesian symmetry that help to attenuate the electric field at the boundaries of the simulation domain and prevent unphysical field reflections from the boundaries. 

We apply the \texttt{Free Tetrahedral} mesh with refined elements in areas of high field concentration and gradients, typically close to the electron's trajectory and nanostructures. We allow for an increase of the size of the mesh elements towards outer boundaries of the simulation domain. The area of PML is meshed by 5-10 \texttt{Swept} layers. The maximal allowed elements' dimensions depend on the simulated energy region and thus on the typical wavelengths involved. We typically use fractions of the typical wavelength for the largest elements.

The electric current density component assigned to either conventional or vortex electron beam is implemented as line \texttt{Edge Current}, whereas the magnetic current is given by line \texttt{Magnetic Current}, as expressed in Eq.~\eqref{Eq:Je}. The (VEB)-EEL probability is evaluated from 3D calculations according to Eq.~\eqref{Eq:eEELS} and Eq.~\eqref{Eq:mEELS} directly with this software using an \texttt{Edge Probe, Integral} along the electron's trajectory between the boundaries at $z_\mathrm{min}$ and $z_\mathrm{max}$. 

All simulations are performed twice: with $\varepsilon(\mathbf{r},\omega)$ corresponding to the probed structure and then with $\varepsilon(\mathbf{r},\omega)=1$ everywhere, so that only the field of the electron is present, preserving the same discretization of the geometrical domains. Afterwards, the loss probability obtained from these two calculations is subtracted to obtain 
only the contribution coming from the induced field arising from 
the interaction of the electron beam with the nanostructure and to correct for the finite length of the electron's trajectory and non-zero values of the fast electron's field at the boundaries of the simulation domain \cite{wiener,raza,ours}.

\section{Electron interacting with a dielectric cylinder}
\label{App:cylinder}
We adapt the expressions from Ref.~\cite{walsh} for the electromagnetic field expressed in cylindrical coordinates $(R,\phi, z)$ for an electron beam moving in vacuum, parallel to an infinite dielectric cylinder along the $z$ axis. The cylinder has a radius $a$ and the electron beam is positioned at radial distance $b>a$ from the center of the cylinder.

The electric and magnetic field components, produced by the fast electron moving in vacuum, in cylindrical coordinates, and Fourier-transformed in the $k_z$ space are:
\begin{align*}
&E_{\mathrm{el},z}(R,\phi,k_z,\omega)=\frac{\mathrm{i}\omega e \delta(\omega/v-k_z)}{v^2 \varepsilon_0\gamma^2}\nonumber\sum\limits_{m=-\infty}^\infty\left[K_m(R\Omega)\nonumber\right.\\
&\left.\times I_m(b\Omega)H(R-b)+K_m(b\Omega)I_m(R\Omega)H(b-R)\right]e^{\mathrm{i}m\phi},\\
&E_{\mathrm{el},R}(R,\phi,k_z,\omega)=\frac{\Omega e \delta(\omega/v-k_z)}{v \varepsilon_0}\sum\limits_{m=-\infty}^\infty\left[K'_m(R\Omega)\right.\nonumber\\
&\left.\times I_m(b\Omega)H(R-b)+K_m(b\Omega)I'_m(R\Omega)H(b-R)\right]e^{\mathrm{i}m\phi},\\
&E_{\mathrm{el},\phi}(R,\phi,k_z,\omega)=\frac{\mathrm{i}\omega e \delta(\omega/v-k_z)}{v \varepsilon_0}\sum\limits_{m=-\infty}^\infty\frac{m}{R}\left[K_m(R\Omega)\right.\nonumber\\
&\left.\times I_m(b\Omega)H(R-b)+K_m(b\Omega)I_m(R\Omega)H(b-R)\right]e^{\mathrm{i}m\phi},\\
&H_{\mathrm{el},z}(R,\phi,k_z,\omega)=0,\\
&H_{\mathrm{el},R}(R,\phi,k_z,\omega)=-\mathrm{i}e\delta(\omega/v-k_z)\sum\limits_{m=-\infty}^\infty\frac{m}{R}\left[K_m(R\Omega)\right.\nonumber\\
&\left. \times I_m(b\Omega)H(R-b)+K_m(b\Omega)I_m(R\Omega)H(b-R)\right]e^{\mathrm{i}m\phi},\\
&H_{\mathrm{el},\phi}(R,\phi,k_z,\omega)=\Omega e\delta(\omega/v-k_z)\sum\limits_{m=-\infty}^\infty\left[K'_m(R\Omega)\right.\nonumber\\
&\left.\times I_m(b\Omega)H(R-b)+K_m(b\Omega)I'_m(R\Omega)H(b-R)\right]e^{\mathrm{i}m\phi},
\end{align*}
 where $I_m(x)$ and $K_m(x)$ are the modified Bessel functions of the first and the second kind, respectively, of order $m$, and $H(x)$ is the Heaviside step function. We also defined $\Omega=\omega/(v\gamma)$. The components of the induced electric and magnetic field inside the cylinder ($R<a$) characterized by a dielectric function $\varepsilon$ are:

\begin{align*}
&E_{\mathrm{in},z}(R,\phi,k_z,\omega)=\sum\limits_{m=-\infty}^\infty-\kappa_i^2a_{\mathrm{e},m} I_m(\kappa_i R)e^{\mathrm{i}m\phi},\\
&E_{\mathrm{in}R}(R,\phi,k_z,\omega)=\sum\limits_{m=-\infty}^\infty\left[\mathrm{i}k_z\kappa_i a_{\mathrm{e},m} I'_m(\kappa_i R)\right.\nonumber\\
&\left.-\frac{\omega\mu_0 m}{R}c_{\mathrm{e},m} I_m(\kappa_i R)\right]e^{\mathrm{i}m\phi},\\
&E_{\mathrm{in},\phi}(R,\phi,k_z,\omega)=\sum\limits_{m=-\infty}^\infty\left[-\frac{m k_z}{R}a_{\mathrm{e},m} I_m(\kappa_i R)\right.\nonumber\\
&\left.-\mathrm{i}\omega \mu_0\kappa_i c_{\mathrm{e},m} I'_m(\kappa_i R)\right]e^{\mathrm{i}m\phi},\\
&H_{\mathrm{in},z}(R,\phi,k_z,\omega)=\sum\limits_{m=-\infty}^\infty-\kappa_i^2 c_{\mathrm{e},m} I_m(\kappa_iR)e^{\mathrm{i}m\phi},\\
&H_{\mathrm{in},R}(R,\phi,k_z,\omega)=\sum\limits_{m=-\infty}^\infty\left[\frac{m\omega\varepsilon\varepsilon_0}{R}a_{\mathrm{e},m} I_m(\kappa_i R)\right.\nonumber\\
&\left. +\mathrm{i}k_z\kappa_i c_{\mathrm{e},m} I'_m(\kappa_i R)\right]e^{\mathrm{i}m\phi},\\
&H_{\mathrm{in},\phi}(R,\phi,k_z,\omega)=\sum\limits_{m=-\infty}^\infty\left[\mathrm{i}\omega\varepsilon\varepsilon_0\kappa_i a_{\mathrm{e},m} I'_m(\kappa_i R)\right.\nonumber\\
&\left. -\frac{m k_z}{R}c_{\mathrm{e},m} I_m(\kappa_i R)\right]e^{\mathrm{i}m\phi},
\end{align*}
whereas the induced electric and magnetic fields 
outside the cylinder ($R>a$) in vacuum are:
\begin{align*}
&E_{\mathrm{out},z}(R,\phi,k_z,\omega)=\sum\limits_{m=-\infty}^\infty-\kappa_o^2 b_{\mathrm{e},m} K_m(\kappa_o R)e^{\mathrm{i}m\phi},\\
&E_{\mathrm{out},R}(R,\phi,k_z,\omega)=\sum\limits_{m=-\infty}^\infty\left[\mathrm{i}k_z\kappa_o b_{\mathrm{e},m} K'_m(\kappa_o R)\right.\nonumber\\
&\left.-\frac{\omega\mu_0 m}{R}d_{\mathrm{e},m} K_m(\kappa_o R)\right]e^{\mathrm{i}m\phi},\\
&E_{\mathrm{out},\phi}(R,\phi,k_z,\omega)=\sum\limits_{m=-\infty}^\infty\left[-\frac{m k_z}{R}b_{\mathrm{e},m} K_m(\kappa_o R)\right.\nonumber\\
&\left.-\mathrm{i}\omega \mu_0\kappa_o d_{\mathrm{e},m} K'_m(\kappa_o R)\right]e^{\mathrm{i}m\phi},\\
&H_{\mathrm{out},z}(R,\phi,k_z,\omega)=\sum\limits_{m=-\infty}^\infty-\kappa_o^2 d_{\mathrm{e},m} K_m(\kappa_oR)e^{\mathrm{i}m\phi},\\
&H_{\mathrm{out},R}(R,\phi,k_z,\omega)=\sum\limits_{m=-\infty}^\infty\left[\frac{m\omega\varepsilon_0}{R}b_{\mathrm{e},m} K_m(\kappa_o R)\right.\nonumber\\
&\left.+\mathrm{i}k_z\kappa_o d_{\mathrm{e},m} K'_m(\kappa_o R)\right]e^{\mathrm{i}m\phi},\\
&H_{\mathrm{out},\phi}(R,\phi,k_z,\omega)=\sum\limits_{m=-\infty}^\infty\left[\mathrm{i}\omega\varepsilon_0\kappa_o b_{\mathrm{e},m} K'_m(\kappa_o R)\right.\nonumber\\
&\left.-\frac{m k_z}{R}d_{\mathrm{e},m} K_m(\kappa_o R)\right]e^{\mathrm{i}m\phi}.
\end{align*}
In the above expressions we assumed only solutions not diverging at $R=0$ and at infinity, and we also defined
$\kappa_\mathrm{i}=\sqrt{k_z^2-\varepsilon\frac{\omega^2}{c^2}}$ and $\kappa_\mathrm{o}=\sqrt{k_z^2-\frac{\omega^2}{c^2}}$. The unknown coefficients $a_{\mathrm{e},m}$, $b_{\mathrm{e},m}$, $c_{\mathrm{e},m}$ and $d_{\mathrm{e},m}$ are obtained by imposing boundary conditions at the boundaries of the cylinder:
\begin{subequations}
\begin{align}
H_{\mathrm{in},z}(R_0,0,q_z,\omega)&=H_{\mathrm{out},z}(R_0,0,q_z,\omega),\\
E_{\mathrm{in},z}(R_0,0,q_z,\omega)&=E_{\mathrm{out},z}(R_0,0,q_z,\omega)\nonumber\\
&+E_{\mathrm{el},z}(R_0,0,q_z,\omega),\\
\varepsilon E_{\mathrm{in},R}(R_0,0,q_z,\omega)&=E_{\mathrm{out},R}(R_0,0,q_z,\omega)\nonumber\\
&+E_{\mathrm{el},R}(R_0,0,q_z,\omega),\\
H_{\mathrm{in},R}(R_0,0,q_z,\omega)&=H_{\mathrm{out},R}(R_0,0,q_z,\omega)\nonumber\\
&+H_{\mathrm{el},R}(R_0,0,q_z,\omega).
\end{align}
\label{Eq:cylinder_coeff}
\end{subequations}

By using the transformation in Eq.~\eqref{Eq:invariance_Maxwell} we can obtain the electric and magnetic field produced in the presence of the cylinder due to the magnetic current component with the unknown coefficients redefined to $a_{\mathrm{m},m}$, $b_{\mathrm{m},m}$, $c_{\mathrm{m},m}$ and $d_{\mathrm{m},m}$, that can be evaluated by applying the same boundary conditions as in Eqs.~\eqref{Eq:cylinder_coeff}. The loss probability per unit trajectory due to the electric and magnetic current is then given by Eq.~\eqref{Eq:cylinder_gam_el} and Eq.~\eqref{Eq:cylinder_gam_mag}, respectively.

\bibliography{VEB_bibliography}

\end{document}